\documentclass[10pt,aps,prx,twocolumn,notitlepage,showpacs,superscriptaddress]{revtex4-1}
\usepackage{amsthm,amsmath,amsfonts,amssymb,verbatim, color}
\usepackage{graphicx}
\usepackage{bm}
\usepackage{epsfig,slashed}
\usepackage[colorlinks=true,citecolor=blue,
linkcolor=blue,urlcolor=blue]{hyperref}
\usepackage{psfrag}

\usepackage{bm}
\usepackage{graphicx}
\usepackage{subfigure}

\newcommand{\sech}{\mathop{\mathrm{sech}}}
\newcommand{\beq}{\begin{equation}}
\newcommand{\eneq}{\end{equation}}
\newcommand{\bs}[1]{\boldsymbol{#1}}
\renewcommand{\b}[1]{{\boldsymbol{#1}}}
\renewcommand{\c}[1]{\mathcal{#1}}
\newcommand{\nn}{\nonumber}

\input{epsf}

\begin{document}

\tolerance 10000

\newcommand{\vk}{{\bf k}}


\title{Evidence 
of a fractional quantum Hall nematic 
phase in a microscopic model} 

\author{N. Regnault}
\affiliation{Department of Physics, Princeton University, Princeton, New Jersey 08544, USA}
\affiliation{Laboratoire Pierre Aigrain, Ecole Normale Sup\'erieure-PSL Research University, CNRS, Universit\'e Pierre et Marie Curie-Sorbonne Universit\'es, Universit\'e Paris Diderot-Sorbonne Paris Cit\'e, 24 rue Lhomond, 75231 Paris Cedex 05, France}

\author{J. Maciejko}
\affiliation{Department of Physics, University of Alberta, Edmonton, Alberta T6G 2E1, Canada}
\affiliation{Theoretical Physics Institute, University of Alberta, Edmonton, Alberta T6G 2E1, Canada}
\affiliation{Canadian Institute for Advanced Research, Toronto, Ontario M5G 1Z8, Canada}

\author{S. A. Kivelson}
\affiliation{Department of Physics, Stanford University, Stanford, California 94305, USA}

\author{S. L. Sondhi}
\affiliation{Department of Physics, Princeton University, Princeton, New Jersey 08544, USA}

\date{\today}

\begin{abstract}
At small momenta, the Girvin-MacDonald-Platzman (GMP) mode in the fractional quantum Hall (FQH) effect can be identified with gapped nematic fluctuations in the isotropic FQH liquid. This correspondence would be exact as the GMP mode softens upon approach to the putative point of a quantum phase transition to a  FQH nematic.  Motivated by these considerations as well as by suggestive evidence of an FQH nematic in tilted field experiments, we have sought evidence of such a nematic FQHE in a microscopic model of interacting electrons in the lowest Landau level at filling factor 1/3. Using a family of anisotropic Laughlin states as trial wave functions, we find a continuous quantum phase transition between the isotropic Laughlin liquid and the FQH nematic. Results of numerical exact diagonalization also suggest that rotational symmetry is spontaneously broken, and that the phase diagram of the model contains both a nematic and a stripe phase.
\end{abstract}

\maketitle

\section{Introduction}


The FQH nematic is a hypothesized state of matter simultaneously characterized by a 
broken symmetry and topological order. In this state, the defining characteristics of topological order (quasiparticles with fractional charge and statistics and topological ground-state degeneracy~\cite{QHE}) coexist with those of nematic order (spontaneous breaking of spatial rotational symmetry~\cite{DeGennes}). Early approaches to broken rotational symmetry in FQH states involved the construction of effective field theories~\cite{balents1996} and trial wave functions~\cite{musaelian1996}. Observations of anisotropic transport in a two-dimensional electron gas (2DEG) at filling fraction $\nu=7/3$~\cite{xia2011} (and, more recently, $\nu=5/2$~\cite{liu2013}) motivated the construction of effective field theories~\cite{mulligan2010,mulligan2011,maciejko2013,you2014} of a putative FQH isotropic-nematic quantum phase transition. 

An independent motivation for the study of such physics comes from the work of Haldane~\cite{haldane2011,haldane2009,haldane2011b}, who proposed a geometrical description of FQH states in which the Girvin-MacDonald-Platzman (GMP) mode~\cite{girvin1985,girvin1986}---a gapped neutral collective mode similar to the phonon-roton mode of superfluids---is interpreted as a fluctuating but unimodular guiding-center spatial metric $g_{ab}(\bs{r},t)$~\cite{luo2016} that describes the shape of the correlation hole in the FQH fluid. In particular, the long-wavelength limit of the GMP mode was identified as a spin-2 (quadrupolar) excitation, analogous to the graviton~\cite{lee1991,dev1992,yang2012b,yang2013b}. 
%
Because a unimodular metric $g_{ab}$ is equivalent via matrix exponentiation $g=\exp Q$ to a traceless symmetric nematic order parameter $Q_{ab}$, the GMP mode of the Laughlin liquid can also be identified as the gapped fluctuations of a nematic order parameter in a disordered (isotropic) phase~\cite{maciejko2013}. 

Deep in the FQH phase, the GMP mode (and all other excitations) occur at high energies, and so are sensitive to microscopic considerations.  However, if 
a transition from an isotropic FQH liquid to a FQH nematic 
can be induced by varying some parameters in the problem, the GMP should soften upon approach to the associated quantum critical point (QCP).  Proximate to the QCP, the physics should be universal and accurately describable by an effective quantum field theory.
This scenario and the resulting effective field theory of the FQH nematic state~\cite{maciejko2013} 
can be realized in a model of 2DEG in a magnetic field with attractive quadrupolar interactions, within the composite-fermion mean-field theory~\cite{you2014}. Ref.~\cite{you2014} also clarified the relation between the Berry phase for nematic fluctuations~\cite{maciejko2013} and the Hall viscosity of the isotropic FQH liquid~\cite{read2009,read2011}.

There are two additional routes to arriving at a FQH nematic in addition to the transition from a gapped FQH state described above. In 
the first the effective field theory approach was extended to the problem of the half-filled Landau level, in order to describe a transition from an isotropic to a nematic composite Fermi liquid~\cite{you2016}. This is relevant in the context of the observation of compressible anisotropic phases in half-filled Landau levels~\cite{lilly1999a,du1999,lilly1999b,pan1999,xia2010,samkharadze2016}. In the composite fermion approach, the FQH nematic is obtained as a Pomeranchuk instability~\cite{pomeranchuk1958,oganesyan2001} of composite fermions~\cite{doan2007}, and the broken rotational symmetry corresponds to the condensation of a fermion bilinear. This is a Fermi-liquid-like or weak-coupling mechanism for the formation of a nematic. In the second, which is a strong-coupling perspective, a nematic electron fluid is viewed as a partially melted solid~\cite{fradkin2010}. In this picture, inspired from the theory of classical liquid crystals, a nematic state is proximate to various phases with broken translational symmetry, such as stripe or smectic phases. If under the influence of some external parameter, topological defects in the latter proliferate in such a way that translational order is melted but orientational order is preserved, a nematic state (or in general, a $\ell$-atic with $\ell\geq 2$) results. This leads, for example, to an alternate description of the compressible nematic phase observed in half-filled Landau levels as a quantum melted stripe phase rather than as a nematic composite Fermi liquid~\cite{fradkin1999,fradkin2000,wexler2001,radzihovsky2002}. According to this perspective, one might expect an incompressible FQH nematic to be proximate not only to an isotropic FQH liquid but also to phases with translational order.

More generally, there is increasingly strong evidence that a host of highly correlated electronic systems support one or another form of electronic liquid crystalline phases~\cite{kivelsonfradkinemery,fradkin2010,vojta2009,fernandes2014,ianfisherscience}. In many ways, the FQH nematic studied here is the paradigmatic example, as the electronic Hamiltonian is simpler, with none of the complexity associated with the solid-state chemistry of typical highly correlated materials, and more symmetric (to good approximation, the system is fully rotationally invariant).  Thus, the results of the present study may be conceptually useful more broadly.

\begin{figure}[t]
\includegraphics[width=0.5\columnwidth]{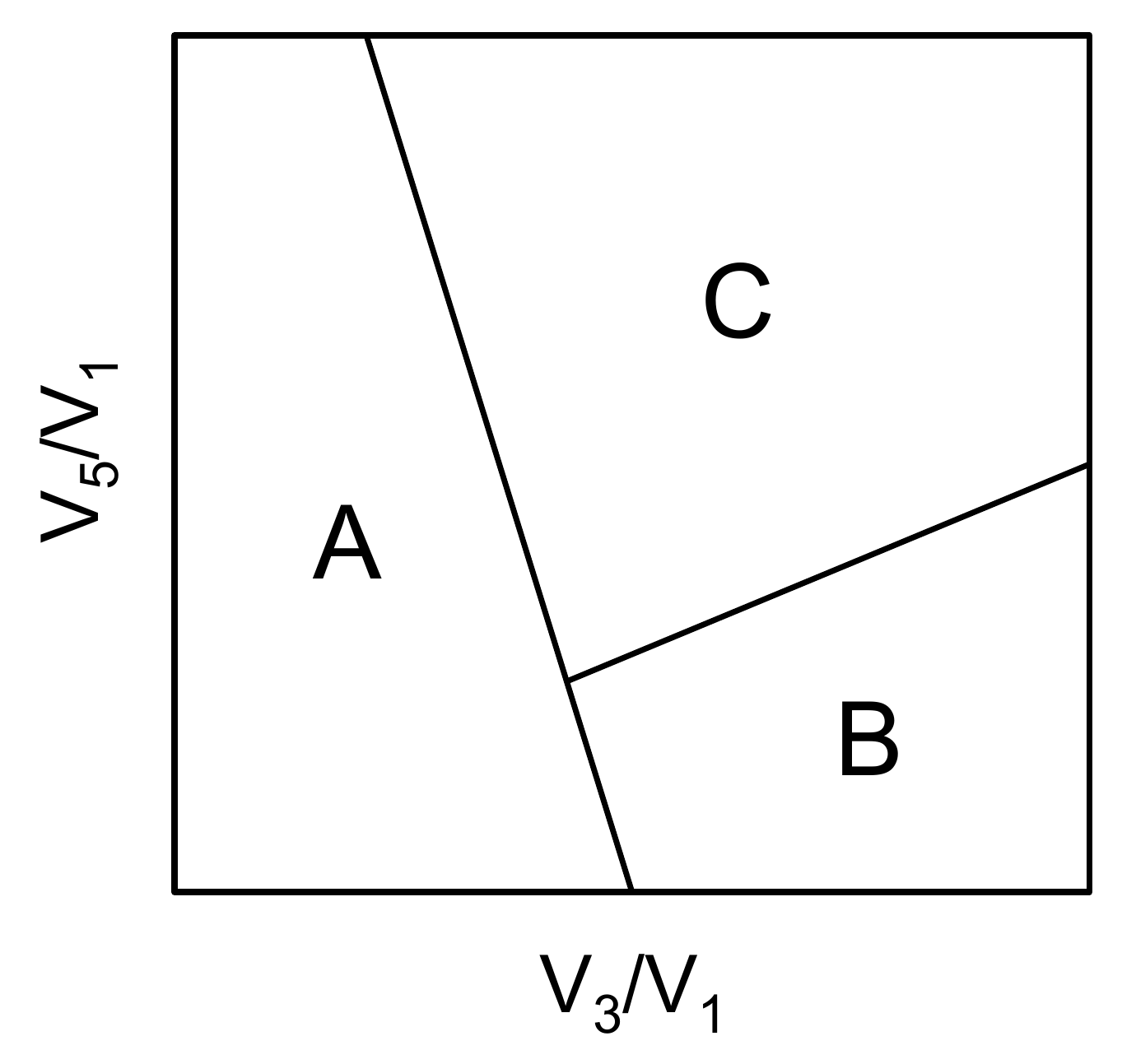}
\caption{Conjectured schematic phase diagram of the $V_1$-$V_3$-$V_5$ model: A = isotropic Laughlin liquid, B = stripe/smectic phase, C = FQH nematic.}\label{fig:fig1}
\end{figure}

In this paper we 
report the results of a microscopic study of spatially ordered phases proximate to the incompressible, isotropic FQH liquid. To do so, we explore possible quantum phase transitions out of the isotropic FQH liquid in a model of interacting electrons in the lowest Landau level (LLL) at filling factor $\nu=1/3$. Our model contains only the first three Haldane pseudopotentials~\cite{haldane1983} $V_1$, $V_3$, and $V_5$, with the ratios $V_3/V_1$ and $V_5/V_1$ as tuning parameters. We study the model with numerical exact diagonalization (ED) for up to $N=13$ electrons as well as with a variational approach using trial wave-functions of a sort that have been used successfully in studies of explicitly anisotropic Hamiltonians~\cite{qiu2012,yang2012,wang2012,papic2013,qiu2013,yang2013,johri2016,apalkov2014,ghazaryan2015,ghazaryan2014}.
 We 
 find that the isotropic $\nu=1/3$ Laughlin liquid is obtained at small values of the tuning parameters but gives way to other phases at larger values (Fig.~\ref{fig:fig1}). 
Our results show (although not entirely unambiguously) the existence of  three phases separated by direct transitions:  (1) An isotropic (Laughlin) FQH liquid;  (2) A nematic FQH liquid crystal; (3) A striped FQH liquid crystal phase (which breaks both rotational and translational symmetry).

\section{Microscopic model}

The general Hamiltonian for a 2DEG in a uniform perpendicular magnetic field and interacting via translationally invariant two-body interactions is
\begin{equation}
H=  \sum_\b{q} V(\b{q}) :\bar{\rho}_\b{q} \bar{\rho}_{-\b{q}}:,\label{GenericHamiltonian}
\end{equation}
when projected to the LLL, which corresponds to ignoring Landau-level mixing effects. In Eq.~(\ref{GenericHamiltonian}), $\bar{\rho}_\b{q}$ is the projected density operator in momentum space and the colons denote normal ordering. The projected two-body interaction $V(\b{q})$ can be decomposed as
\begin{equation}
V(\b{q})=\sum_{n=1,3,5,\ldots} V_n {\cal L}_n(\b{q}^2)\label{Pseudopotentials},
\end{equation}
where $V_n$ are the Haldane pseudopotentials and ${\cal L}_n$ is the $n$th Laguerre polynomial. The model we consider is one in which $V_1,V_3,V_5$ are nonzero and all the other pseudopotentials are set to zero. Thus our Hamiltonian depends on two tuning parameters $V_3/V_1$ and $V_5/V_1$,
\begin{eqnarray}
H(V_3/V_1,V_5/V_1)&=&V_1\sum_\b{q}\biggl(\c{L}_1(\b{q}^2)+\frac{V_3}{V_1}\c{L}_3(\b{q}^2)\nn\\
&&+\frac{V_5}{V_1}\c{L}_5(\b{q}^2)\biggr):\bar{\rho}_\b{q} \bar{\rho}_{-\b{q}}:,\label{v1v3v5}
\end{eqnarray}
with $V_1$ simply setting the overall energy scale. For $V_3/V_1=V_5/V_1=0$, one recovers the model Hamiltonian for which the Laughlin state is the exact zero-energy ground state at $\nu=1/3$. For our calculations, we consider $N$ interacting electrons in the LLL on a torus geometry pierced by $N_\Phi$ flux quanta. The torus is spanned by the two vectors $L_x \b{e}_x$ and $L_y \b{e}_y$, with two orthogonal unit vectors $\b{e}_x$ and $\b{e}_y$. The aspect ratio of the torus is defined as $L_y/L_x = 1 + \delta$. For $\delta=0$, the system has a square aspect ratio and exhibits an additional $C_4$ symmetry. We fix the filling factor $\nu=N/N_\Phi$ to be $\nu=1/3$, and neglect spin effects.

\section{Variational approach}
\label{sec:variational}

To search for a possible isotropic-to-nematic transition in the microscopic model (\ref{v1v3v5}), a natural first line of attack is a variational approach. This requires (1) a family of trial wave functions parametrized by an appropriate set of variational parameters, and (2) a microscopic definition of an order parameter that can be calculated from these wave functions.

In 2D systems such as the FQH liquids of interest to us, nematic order is described in the long-wavelength limit by a real traceless symmetric rank-two tensor $Q_{ab}$ with two independent parameters $Q_{11}=-Q_{22}$ and $Q_{12}=Q_{21}$~\cite{DeGennes}. It is convenient to use a complex representation $Q\equiv Q_{11}+iQ_{12}=|Q|e^{i\phi}$ where $|Q|$ and $\phi$ are the amplitude and phase of the nematic order parameter, respectively~\cite{fradkin2010b}. In this representation, phase rotations are spatial rotations; here we define $\phi$ such that a rotation of it by $2\pi$ corresponds to a physical spatial rotation by $\pi$, which leaves the nematic order parameter invariant. In an isotropic phase, $|Q|$ vanishes and $\phi$ is free to fluctuate, while in a nematic phase, $|Q|$ is nonzero and $\phi$ selects a specific direction, breaking rotational symmetry spontaneously.

Building on the equivalence between Haldane's unimodular metric $g_{ab}$ and a nematic order parameter~\cite{maciejko2013},
\begin{align}\label{gab}
g_{ab}=\left(\begin{array}{cc}
\mathop{\mathrm{ch}}|Q|+\mathop{\mathrm{sh}}|Q|\cos\varphi & \mathop{\mathrm{sh}}|Q|\sin\varphi \\
\mathop{\mathrm{sh}}|Q|\sin\varphi & \mathop{\mathrm{ch}}|Q|-\mathop{\mathrm{sh}}|Q|\cos\varphi
\end{array}\right),
\end{align}
we use the anisotropic LLL wave functions of Ref.~\cite{qiu2012} as a continuous family of trial ground-state wave functions $|\Psi(|Q|,\phi)\rangle$ parametrized by $|Q|$ and $\phi$ (see Appendix~\ref{app:trialWF} for the explicit form of these wave functions). To determine the possibility of an FQH nematic phase in the Hamiltonian~(\ref{v1v3v5}), we minimize the variational energy
\begin{align}\label{VarE}
&E(|Q|;V_3/V_1,V_5/V_1)=\nonumber\\
&\hspace{10mm}\frac{\langle\Psi(|Q|,\phi)|H(V_3/V_1,V_5/V_1)|\Psi(|Q|,\phi)\rangle}
{\langle\Psi(|Q|,\phi)|\Psi(|Q|,\phi)\rangle},
\end{align}
with respect to the variational parameter $|Q|$. Unlike in previous studies~\cite{qiu2012,yang2012,wang2012,papic2013,qiu2013,yang2013,johri2016} where the optimal metric $g_{ab}$ is determined variationally for a system with an anisotropic band effective mass or an anisotropic dielectric tensor, i.e., for a microscopic Hamiltonian that \emph{explicitly} breaks rotation symmetry, here our microscopic Hamiltonian (\ref{v1v3v5}) is rotationally invariant and the variational energy (\ref{VarE}) is independent of $\phi$. However, rotation symmetry can be broken \emph{spontaneously} if Eq.~(\ref{VarE}) is minimized by a nonzero nematic amplitude $|Q|$ for certain values of $V_3/V_1$ and $V_5/V_1$.

Finally, we must find a definition of the nematic order parameter in terms of the electronic degrees of freedom of our microscopic model. In principle, any microscopic observable with quadrupolar symmetry is a suitable candidate. Consider for example the quantity
\begin{align}\label{fd}
f(\b{d})=\langle \Psi|\bar{\rho}(\b{r}+\b{d}/2)\bar{\rho}(\b{r}-\b{d}/2)|\Psi\rangle,
\end{align}
where $\b{r}$ is the position vector, $\bar{\rho}(\b{r})$ is the projected density operator in real space, $|\Psi\rangle$ is the ground state, and $\b{d}$ is a vector with a microscopic length (so the order parameter remains local) and angle $\theta$ in the plane. The length of $\b{d}$ only affects the overall scale of the order parameter amplitude and can be chosen for simplicity to be the magnetic length $\ell_B$, which is the only length scale in the problem. For a translationally invariant ground state, Eq.~(\ref{fd}) is independent of $\b{r}$. For a rotationally invariant ground state, $f(\b{d})=0$. Furthermore, $f(\b{d})=f(-\b{d})$, hence $f(\b{d})$ measures the breaking of rotational symmetry up to the equivalence $\b{d}\sim-\b{d}$, as does a nematic order parameter (which can also be thought of as a headless vector~\cite{DeGennes}). Expanding $f(\b{d})$ in angular momentum components, we find that the first nontrivial term has angular momentum $l=2$ and thus its complex coefficient $f_2=\int_0^{2\pi}\frac{d\theta}{2\pi}e^{-2i\theta}f(\b{d})$ can be used as a microscopic definition of our complex nematic order parameter. Because our numerical simulations are performed on a torus, the full $SO(2)$ rotation symmetry is in fact explicitly broken to a discrete $C_4$ rotation symmetry by the periodic boundary conditions (see Appendix~\ref{app:finitesize} for a detailed discussion). The resulting Ising nematic~\cite{abanin2010,metlitski2010} is thus more appropriately described by an order parameter $N_{x^2-y^2}\propto f(\hat{x})-f(\hat{y})$ with $d_{x^2-y^2}$ symmetry,
\begin{align}\label{NematicOP}
N_{x^2-y^2}\equiv\sum_\b{q}(\cos q_x-\cos q_y)\langle \Psi|\bar{\rho}_{\b{q}}\bar{\rho}_{-\b{q}}|\Psi\rangle,
\end{align}
working in units such that $\ell_B=1$, and assuming nematic order along the $x$ axis without loss of generality.

\begin{figure}[t]
\includegraphics[width=\columnwidth]{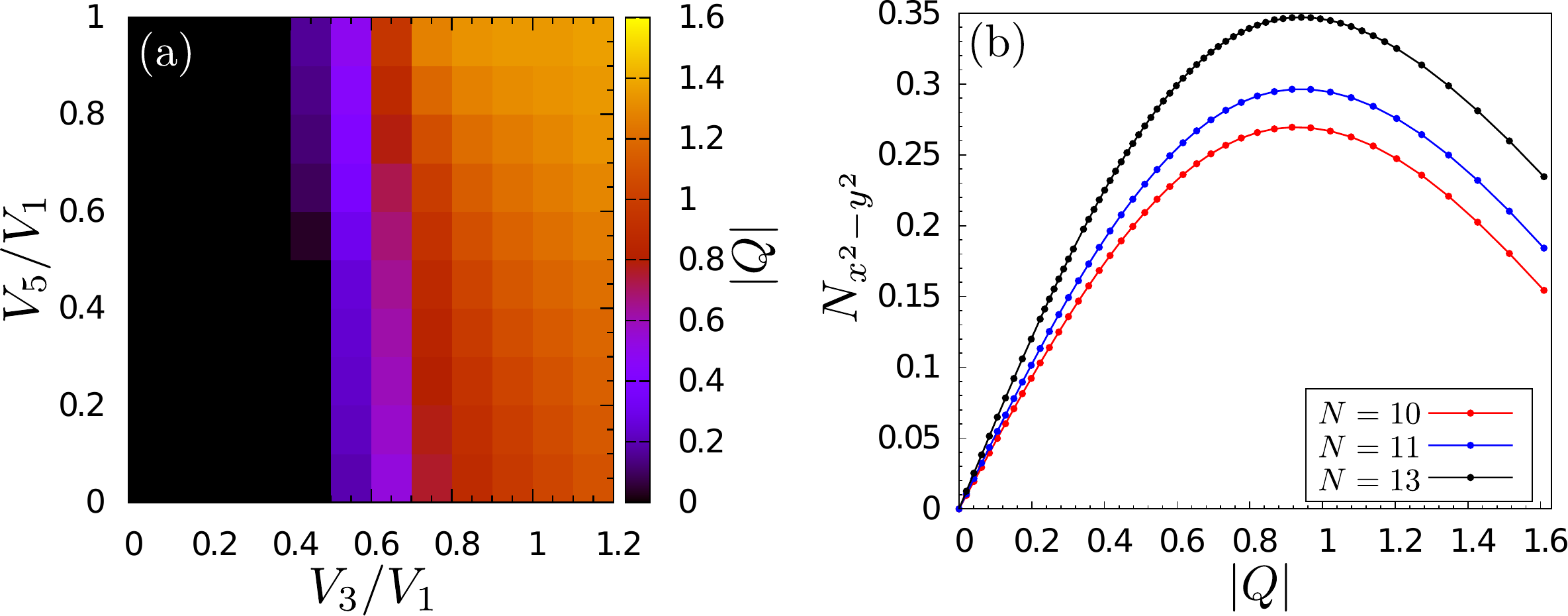}
 \caption{(a) Value of the variational parameter $|Q|$ that minimizes the ground-state energy (\ref{VarE}) for a system with $N=13$ electrons. A continuous quantum phase transition from an isotropic to a nematic FQH state is seen at $V_3/V_1\approx 0.6$. (b) Amplitude of the microscopic nematic order parameter (\ref{NematicOP}) as a function of $|Q|$.}
  \label{fig:fig6}
\end{figure}

In Fig.~\ref{fig:fig6}(a) we plot the value of the variational parameter $|Q|$ that minimizes the variational energy (\ref{VarE}), while in Fig.~\ref{fig:fig6}(b) we give the correspondence between $|Q|$ and the microscopic nematic order parameter $N_{x^2-y^2}$, i.e., Eq.~(\ref{NematicOP}) evaluated in the optimal trial state. The isotropic FQH liquid is stable up to a value $V_3/V_1\approx 0.6$, beyond which rotation symmetry is broken spontaneously. Given that the trial states $|\Psi(|Q|,\phi)\rangle$ have the same topological order as the isotropic $\nu=1/3$ Laughlin state for all $|Q|$ and $\phi$~\cite{qiu2012}, this broken-symmetry state is a FQH nematic. In this variational approach, the isotropic-nematic transition is continuous; however, in other approaches such as composite fermion mean-field theory~\cite{you2014} the transition is sometimes found to be first order. Figure~\ref{fig:fig6} was calculated numerically and is shown for $N=13$, but finite-size effects are negligible given the variational nature of this approach. The variational approach thus suggests a first phase boundary in our schematic phase diagram (Fig.~\ref{fig:fig1}), i.e., the nearly vertical phase boundary that signals the destruction of the isotropic Laughlin liquid (phase A).

\section{Exact diagonalization}\label{sec:ED}

\begin{figure}[t]
\includegraphics[width=\columnwidth]{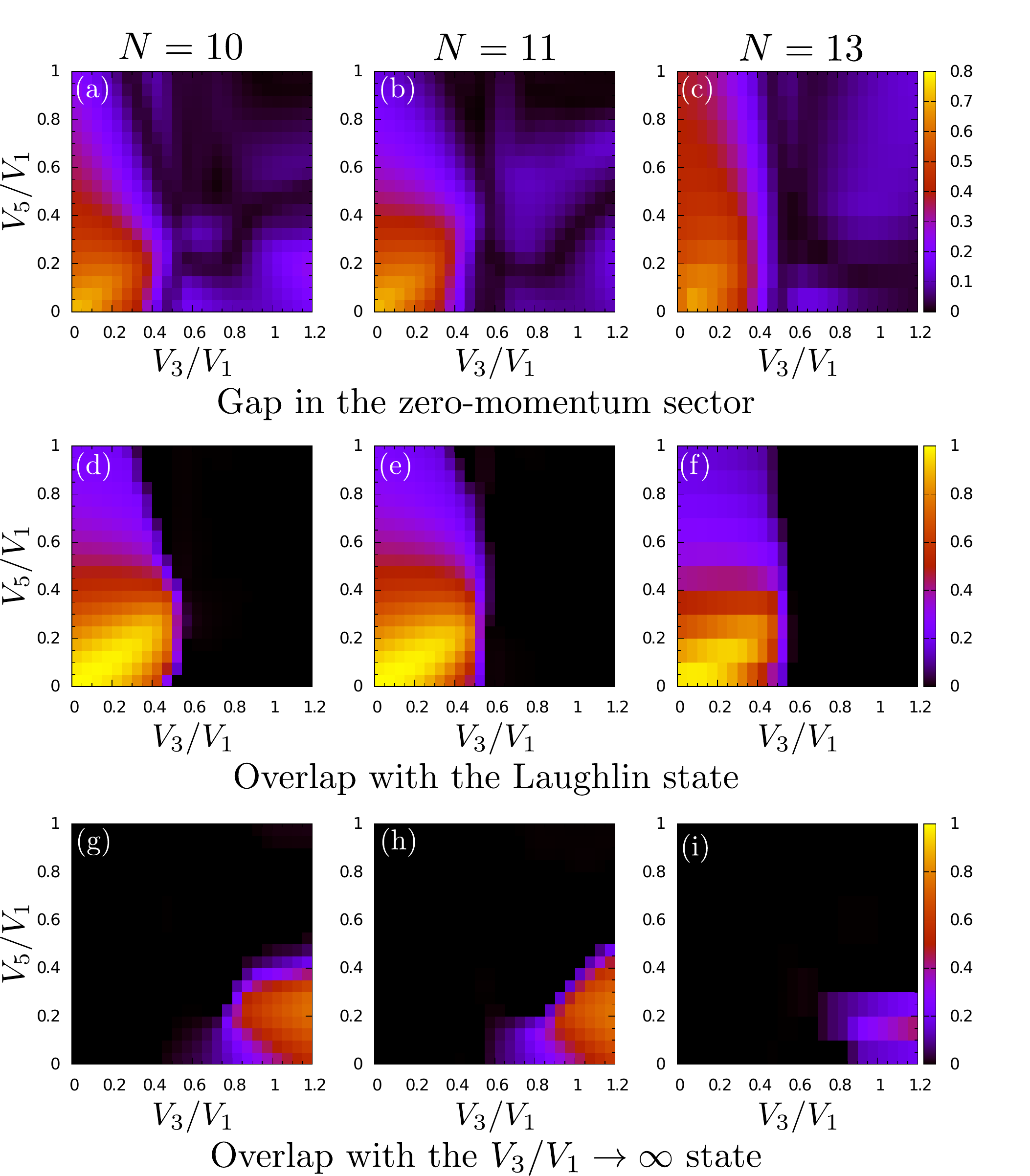}
\caption{Exact diagonalization study of the $V_1$-$V_3$-$V_5$ model with $N$ electrons: (a)-(c) gap between ground and first excited state in the zero-momentum sector; (d)-(f) overlap of the ground state with the $\nu=1/3$ Laughlin state; (g)-(i) overlap of the ground state with the $V_3/V_1\rightarrow\infty$ ground state.}\label{fig:fig2}
\end{figure}

To complement our variational approach, we study the model (\ref{v1v3v5}) with ED for up to $N=13$ electrons. Our main results are shown in Fig.~\ref{fig:fig2}. In order to compare gaps between different values of $V_3/V_1$, we have set the energy scale to be one for the two-particle problem, irrespective of the value of $V_3/V_1$. While finite-size effects are still noticeable from a quantitative standpoint, from a qualitative standpoint the combined data of the gap in the zero-momentum sector [Fig.~\ref{fig:fig2}(a)-(c)] and the overlap of the ground state with the Laughlin [Fig.~\ref{fig:fig2}(d)-(f)] and $V_3/V_1\rightarrow\infty$ [Fig.~\ref{fig:fig2}(g)-(i)] ground states allow us to further refine our schematic phase diagram (Fig.~\ref{fig:fig1}). Three phases are noticeable, separated by gap-closing transitions. In agreement with the variational results, phase A occurs for $V_3/V_1$ less than $\sim 0.5$ and appears to persist for relatively large $V_5/V_1$. It is gapped and adiabatically connected to the isotropic $\nu=1/3$ Laughlin liquid. The second phase (denoted by B) occurs for $V_3/V_1\gtrsim 0.5$ and small $V_5/V_1$. The gap in the zero-momentum sector is much smaller than that of the Laughlin state and appears to be strongly size-dependent. The ground state in this part of the phase diagram has significant overlap with the $V_3/V_1\rightarrow\infty$ ground state (obtained by considering Eq.~(\ref{Pseudopotentials}) with $V_3$ nonzero only). Since this overlap grows with increasing $V_3/V_1$, we suspect the ground state is adiabatically connected to the $V_3/V_1\rightarrow\infty$ ground state. This picture holds when looking at the gap to the first excited state irrespective of its momentum (such a calculation can only be performed for $N=10$ and $N=11$). The $V_3$-only model has only been the object of a few studies~\cite{Wojs-PhysRevB.69.205322,Wojs-PhysRevB.71.245331,Mukherjee-PhysRevLett.112.016801}, and we discuss it further in Sec.~\ref{sec:largeV3}. Finally, the third phase (denoted by C) occurs for large $V_3/V_1$ and $V_5/V_1$. It has negligible overlap with both the Laughlin state and the $V_3/V_1\rightarrow\infty$ ground state; we discuss it further in Sec.~\ref{sec:largeV3V5}. Note that we did not consider a system with $N=12$ electrons. Indeed, the Laughlin liquid has an enhanced stability due to commensuration effects (see Appendix~\ref{app:numerical}).

As already mentioned, the gap-closing phase boundary obtained in ED between phase A and phases B and C roughly matches the isotropic-nematic phase boundary obtained in the variational approach. Since phases B and C are on the nematic side of the transition, this suggests they are characterized by some form of spatial order in the thermodynamic limit. As we discuss in Sec.~\ref{sec:largeV3V5}, phase C exhibits a large nematic susceptibility, and we conjecture it is an FQH nematic in the thermodynamic limit. Given that the variational approach with a single nematic variational parameter predicts an FQH nematic also in the region corresponding to phase B, while our ED results suggest this is a distinct phase but also with spatial order (as we will argue in Sec.~\ref{sec:largeV3}), we conjecture that phase B is a spatially ordered phase proximate to a nematic, that has a lower energy than the nematic in this region of parameter space. The most natural possibility is a phase with spontaneously broken translation symmetry, i.e., a smectic/stripe phase, from which the FQH nematic descends by quantum melting at the B-C phase boundary. Such phases have been considered in the quantum Hall context before~\cite{fradkin1999,fradkin2000,wexler2001,radzihovsky2002}.

\subsection{Large $V_3$: possible stripe phase}\label{sec:largeV3}

\begin{figure}[t]
\includegraphics[width=0.9\columnwidth]{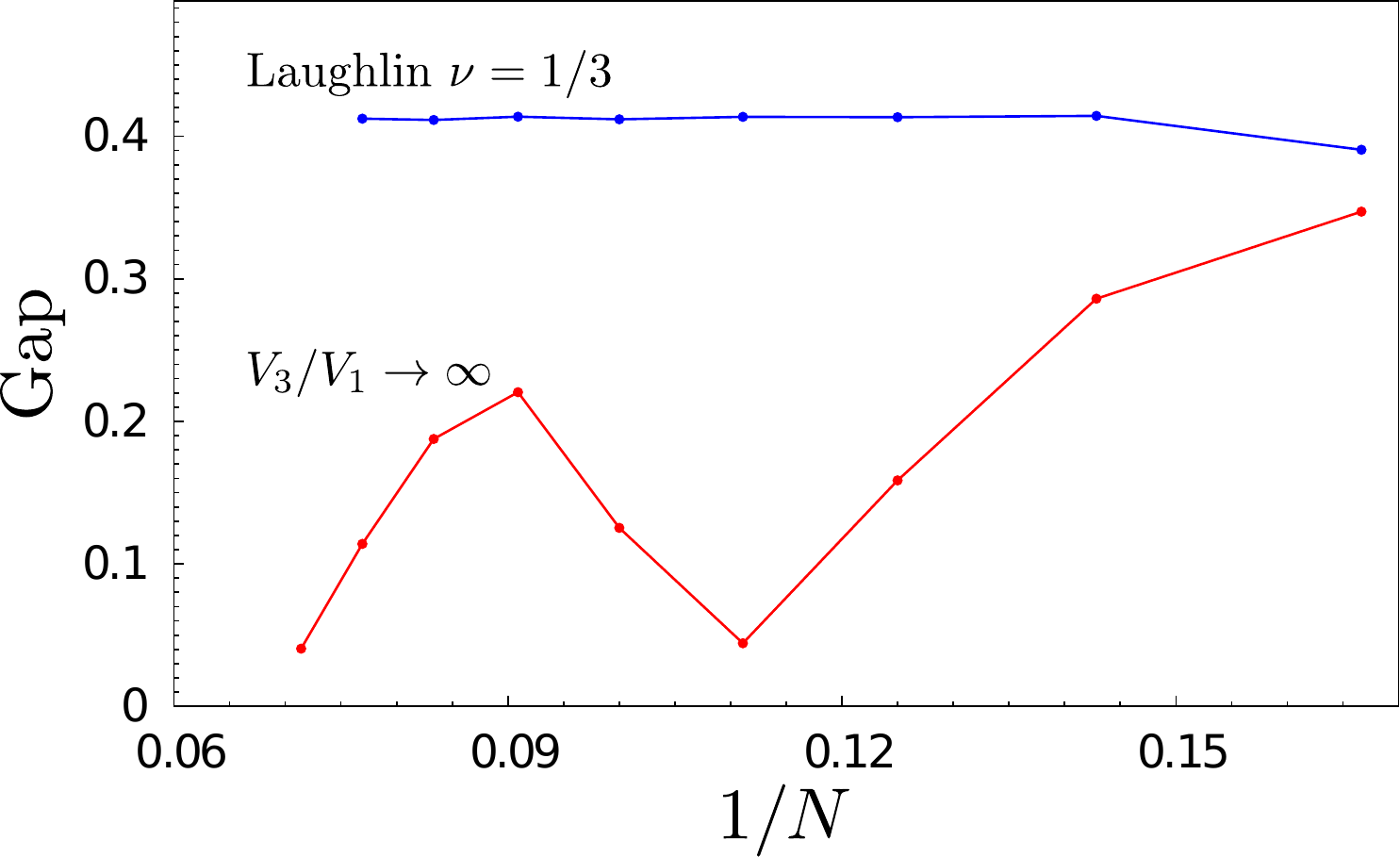}
\caption{Dependence of gap to first excited state on system size: comparison between the Laughlin liquid (blue curve, $V_1=1$ and $V_3=V_5=0$) and the pure $V_3$ phase (red curve, $V_3=1$ and $V_1=V_5=0$ i.e. $V_3/V_1\rightarrow\infty$). The gap was computed up to $N=14$ (resp. $N=13$) electrons for the pure $V_3$ phase (resp. Laughlin liquid).}\label{fig:fig3}
\end{figure}

Compared to the $V_1$-only Hamiltonian whose exact ground state at filling factor $\nu=1/3$ is the celebrated Laughlin state, the $V_3$-only Hamiltonian at the same filling factor has been studied much less (the exact ground state of the $V_3$-only Hamiltonian is the $\nu=1/5$ Laughlin state, but at filling factor $1/5$). A series of numerical ED studies~\cite{Wojs-PhysRevB.69.205322,Wojs-PhysRevB.71.245331,Mukherjee-PhysRevLett.112.016801} have argued that the ground state in this limit is an incompressible state with shift 7 on the sphere and topologically distinct from the Laughlin state, the so-called W\'{o}js-Yi-Quinn (WYQ) state. Interpreting the $V_3$-only Hamiltonian as describing the interaction of composite fermions in the second LL at filling factor $\nu_\text{CF}=1+1/3$, the resulting WYQ state of composite fermions was proposed as a candidate state for a recently discovered FQHE at filling factor $\nu=4/11$~\cite{pan2015,samkharadze2015}.

Here we propose the alternate possibility that the $V_3/V_1\rightarrow\infty$ ground state, and by adiabatic continuity, phase B in Fig.~\ref{fig:fig1}, might be gapless in the thermodynamic limit. First, as mentioned before, the gap in phase B appears to be strongly size-dependent. In Fig.~\ref{fig:fig3} we compare the dependence on system size $N$ of the gap between ground and first excited state for the Laughlin liquid ($V_1$-only model) and the $V_3$-only model. While the gap in the Laughlin liquid is essentially system-size independent, the gap in the pure $V_3$ model exhibits a strong dependence on $N$ (and almost closes at $N=14$ electrons, the largest size we can reach), which suggests the existence of gapless degrees of freedom in the thermodynamic limit. It is interesting to note in this context that in a recent study~\cite{mukherjee2015}, the neutral collective mode spectrum of the composite fermion WYQ state mentioned above was found to be nearly gapless, with the wave vector and energy of the magnetoroton minimum an order of magnitude smaller than in the $\nu=1/3$ Laughlin state.

\begin{figure}[t]
\includegraphics[width=\columnwidth]{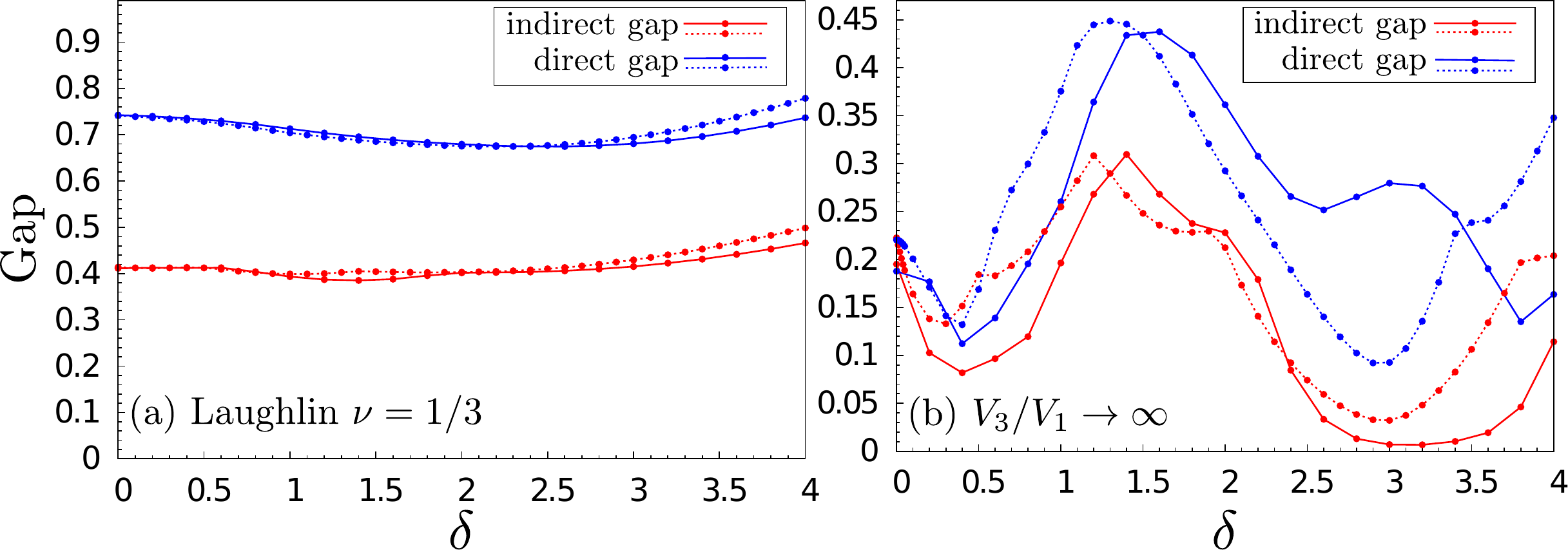}
\caption{Dependence of gap to first excited state on aspect ratio $\delta=L_y/L_x-1$ of the torus in both the zero-momentum sector (blue lines) and at finite momentum (red lines). We have considered both $N=12$ electrons (solid lines) and $N=11$ electrons (dotted lines): (a) Laughlin liquid ($V_3=V_5=0$), (b) pure $V_3$ phase ($V_3/V_1\rightarrow\infty$).}\label{fig:fig4}
\end{figure}

Second, in Fig.~\ref{fig:fig4} we plot the dependence of the gap on the aspect ratio $\delta$ of the torus. Here we plot both the direct gap (gap in the zero-momentum sector) and the absolute gap, which in these systems is indirect (i.e., occurs at finite momentum). In the Laughlin liquid there is essentially no dependence of the gaps on the aspect ratio, as one expects for an isotropic incompressible phase. By contrast, in the pure-$V_3$ case both direct and indirect gaps are strongly dependent on the aspect ratio, which is at odds with the scenario proposed previously of an isotropic, incompressible topological phase. Within the limitations of numerical ED on small systems, this result nonetheless suggests that the ground state of the $V_3$-model is strongly susceptible to spatial perturbations that break the spatial $C_4$ rotational symmetry of the torus. In the thermodynamic limit, this is suggestive of a phase with a spontaneously broken spatial symmetry. As we discuss in Sec.~\ref{sec:largeV3V5}, phase C is characterized by a large nematic susceptibility, leading us to speculate that it becomes an FQH nematic in the thermodynamic limit. Since phase B is a distinct phase, but proximate to a nematic, we conjecture it is a smectic/stripe phase.

\subsection{Large $V_3$ and $V_5$: possible FQH nematic}\label{sec:largeV3V5}

In the large $V_3,V_5$ regime, the ground state has essentially zero overlap with both the Laughlin state and the $V_3/V_1\rightarrow\infty$ ground state; we surmise it corresponds to a distinct phase of matter in the thermodynamic limit (phase C in Fig.~\ref{fig:fig1}). As in the case of phase B, the gap in phase C appears to be strongly size-dependent [Fig.~\ref{fig:fig2}(a)-(c)], suggestive of a gapless phase in the thermodynamic limit. Combined with the variational results of Sec.~\ref{sec:variational}, this strongly suggests phase C is a FQH nematic.

To further investigate this possibility and go beyond the variational approach, which is necessarily biased, we would like to show directly that the ground state obtained in ED in the large $V_3,V_5$ regime exhibits nematic order. However, since there is no spontaneous symmetry breaking in a finite system, we can only calculate a nematic susceptibility, defined as the rate of change of the nematic order parameter with respect to a suitable symmetry-breaking field. Since a nonzero aspect ratio parameter $\delta$ explicitly breaks the $C_4$ rotation symmetry of the torus, we adopt the following microscopic definition of nematic susceptibility,
\begin{align}\label{chiQ}
\chi_Q\equiv\lim_{\delta\rightarrow 0}\frac{\partial N_{x^2-y^2}(\delta)}{\partial\delta},
\end{align}
where the nematic order parameter (\ref{NematicOP}) is calculated numerically with the ED ground state $|\Psi\rangle$ (by opposition to the trial ground state) and as a function of $\delta$. In practice, we obtain $\chi_Q$ by numerically determining the slope of the curve of $N_{x^2-y^2}(\delta)$ vs $\delta$ for small values of $\delta$.

\begin{figure}[t]
\includegraphics[width=\columnwidth]{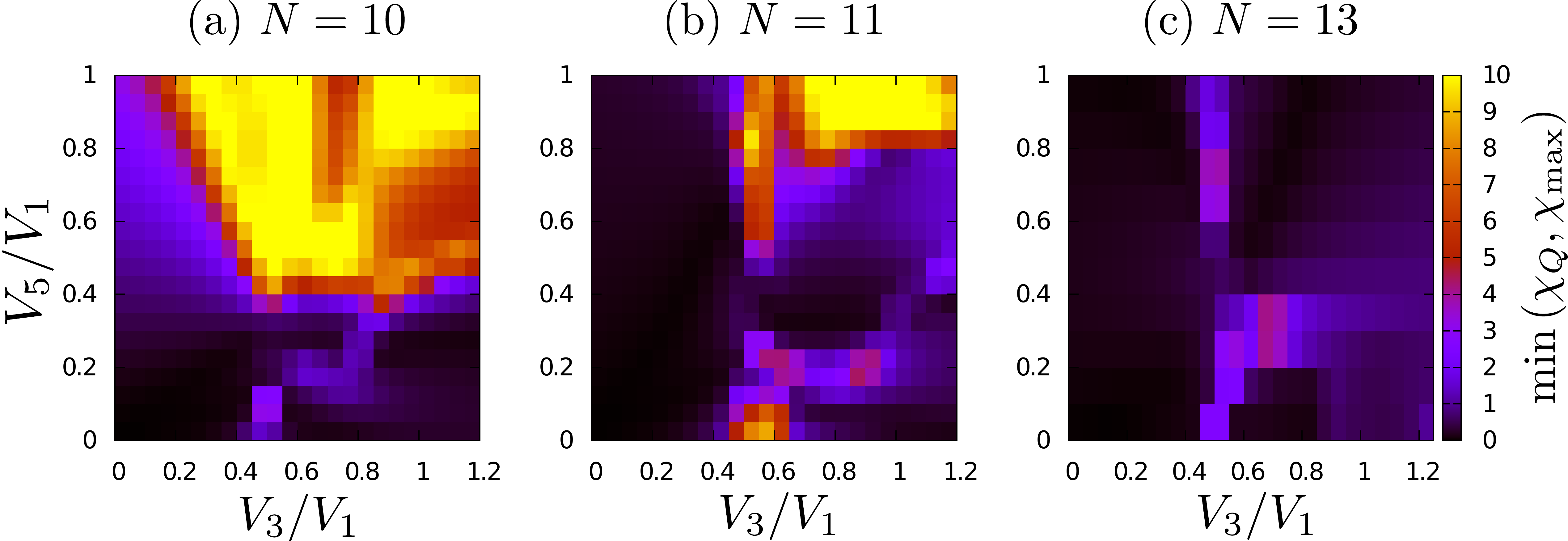}
\caption{Ground-state nematic susceptibility (\ref{chiQ}), capped for readability at $\chi_\textrm{max}=10$.}\label{fig:fig5}
\end{figure}

In Fig.~\ref{fig:fig5} we plot the nematic susceptibility $\chi_Q$ as a function of the tuning parameters $V_3/V_1$ and $V_5/V_1$, for different system sizes. While we are clearly in a regime where finite-size effects are still important, at least for $N=10$ and $N=11$ electrons there is a strong nematic susceptibility in this phase. This is evidence of strong nematic correlations, and further supports our conjecture that phase C is an FQH nematic in the thermodynamic limit.

\section{Implications for experiments}

Experimental evidence of nematic $\nu=7/3$ and $\nu=8/3$ FQH states was reported in the tilted field experiments of Xia {\it et al.}~\cite{xia2011}.  They argued convincingly that the tilted field has two important effects in the particular devices they studied:  Of course, it produces a symmetry breaking field in the 2DEG which defines a preferred nematic ``x'' axis parallel to the in-plane component of the magnetic field.  
It also affects the effective interactions, and it is apparently this latter effect that tunes the system through an isotropic-to-nematic transition.  In the relevant range of tilt angles, the Hall conductance is appropriately quantized at low temperatures.  However, the longitudinal resistivity exhibits an astonishing $T$ dependence: 
 While at elevated $T$, $\rho_{xx}$ and $\rho_{yy}$ differ by a tilt-angle dependent factor,  both decrease with decreasing $T$ in a similar fashion.  However, below a critical temperature $T_\text{nem} \sim 50$~mK, $\rho_{xx}$ exhibits a sharp change such that it is an increasing function of decreasing $T$, while $\rho_{yy}$ continues its smooth decrease, tending  to 0 as $T\to 0$.  

The existence of a finite-$T$  transition is expected~\cite{fradkin2000} for a nematic phase, and this and the large enhancement of the resistance anisotropy are the strongest pieces of evidence in favor of this identification~\cite{fradkin2000,mulligan2010,mulligan2011}.  However, it is far from obvious how to understand the particulars of the results.  Most obviously, the quasiparticle creation energy and character should not change dramatically across the nematic transition, and consequently well inside the nematic state, either in the regime of thermally activated transport or variable range hopping, $\rho_{xx}/\rho_{yy}$ should approach a $T$ independent constant value, which is not readily reconciled with the strongly $T$ dependent anisotropy that is observed.

To resolve this issue, we offer a speculative proposal, which could be readily tested in future experiments.  One of the most important characteristics of classical liquid crystals is that in appropriate circumstances they form macroscopic textures --- textures whose length scales are set by the size of system.  The ubiquitous existence of such textures reflects the fact that the broken symmetries involved are spatial as opposed to internal symmetries.  The nature of the textures depends sensitively on the nature of the broken symmetries.  For theory at the level of the effective  Hamiltonian studied in the present paper, there is full $U(1)$ rotational symmetry, so the nematic is an XY order parameter.  However, in the GaAs heterostructures in which the 2DEG is realized, corrections to the effective mass approximation break that symmetry down to a $C_4$ discrete subgroup.  These corrections are small in proportion to $(a/\ell_B)^4$ (see Appendix~\ref{app:anisotropy}), so they are unimportant for most purposes, but at long distances in the broken symmetry phase they necessarily play an essential role, and correspondingly any macroscopic textures should be characteristic of an Ising order parameter.  

In the present context, we note that generically one particular order parameter orientation is favored at any edge of the system. 
So long as this preference is sufficiently strong, it must be satisfied in the equilibrium state, which forces the order parameter to be spatially varying in the bulk.  In the case of an Ising order parameter and a square sample in the presence of a weak symmetry breaking field, or more generally in any rectangular sample,  this forces the existence of two domain walls which split the sample into three macroscopic domains running along one direction.  The nature of the states associated with the domain wall is, itself, an interesting problem whose solution depends on a number of microscopic considerations;  however, typically such domain walls are associated with a dramatically reduced gap to current carrying excitations.  Where this is the case, the existence of such textures below $T_\text{nem}$ could account for the opposite $T$ dependence of $\rho_{xx}$ and the suddenness of its onset, even in the presence of a  symmetry breaking field.  Clearly, further theoretical and experimental work is necessary to test this proposal.

Assuming that what is observed in these experiments is, indeed, an FQH nematic, there are a number of other interesting quantities that could be measured.  Most importantly from the present viewpoint, the GMP mode, which can be observed as a sharp mode in Raman~\cite{golkar2016} or acoustic wave~\cite{yang2016} spectroscopy, should soften as the transition is approached from the isotropic side.  Observing this mode would confirm its identification as the soft-mode associated with a proximate nematic state;  more generally, it would be interesting to observe as the condensed matter version of a graviton.

\section{Conclusion}

In conclusion, we have explored the possibility of engineering quantum phase transitions out of the isotropic FQH liquid towards spatially ordered phases in a microscopic model of an interacting 2DEG in a perpendicular magnetic field. Combining a variational study with ED results for up to $N=13$ electrons, we arrived at the schematic phase diagram illustrated in Fig.~\ref{fig:fig1}. Phase A was seen to be adiabatically connected to the isotropic $\nu=1/3$ Laughlin state, and thus represents the region of stability of the isotropic FQH liquid. In phase B, numerical evidence suggests that this phase is adiabatically connected to the $V_3/V_1\rightarrow\infty$ ground state. The many-body gap of the latter was seen to be strongly dependent on system size and aspect ratio of the torus on which our simulations were defined, which suggests that phase B would have spontaneously broken spatial symmetries in the thermodynamic limit. Finally, a variational analysis with anisotropic Laughlin trial wave functions combined with the study of a suitably defined nematic susceptibility led us to propose that phase C could be a FQH nematic in the thermodynamic limit. General arguments suggest that a nematic should be proximate to a phase with translational order: we were thus led to conjecture that phase B could be a smectic/stripe phase, which melts into a nematic at the B-C phase boundary.

Of course, the main limitation of our study is the neglect of quantum fluctuations in our variational analysis as well as the presence of finite-size effects (especially in phases B and C). We hope nonetheless that our work will stimulate further numerical studies of the model Hamiltonian considered here. In particular, it would be interesting to study this model and test the conjectures presented here using algorithms recently developed to apply the density-matrix renormalization group to FQH systems~\cite{zaletel2012,zaletel2013,zaletel2015}, which might reduce finite-size effects.

Finally we note that weak disorder will destroy both the nematic and the stripe phase in $d=2$ but the nematic will have a quadratically
longer correlation length in this limit and exponentially longer if the spatial anisotropy is taken into account. So the practical effect of weak
disorder is to stabilize the nematic relative to the stripe phase~\footnote{For example, see Ref.~\cite{nie2014}.}.

\acknowledgements

We wish to acknowledge helpful discussions with E. Fradkin and Z. Papi\'{c}. N.~R. was supported by the Packard Foundation, MURI-130-6082, a Keck grant and the Princeton Global Scholarship. J.~M. was supported by NSERC grant \#RGPIN-2014-4608, the CRC Program, CIFAR, and the University of Alberta. S.~A.~K. was supported by NSF-DMR via Grant No.~1265593, and S.~L.~S. was supported by NSF-DMR via Grant No.~1311781.

\appendix

\section{Variational wave functions for the FQH isotropic-to-nematic transition}
\label{app:trialWF}

In Sec.~\ref{sec:variational} we use the anisotropic Laughlin states $\Psi_L^{1/3}(\gamma)$ at filling factor $\nu=1/3$ constructed in Ref.~\cite{qiu2012} as a continuous family of trial ground-state wave functions for the FQH isotropic-to-nematic transition. These anisotropic Laughlin states are many-body wave functions at filling factor $1/q$ in the LLL,
\begin{align}\label{AnisLaughlin}
\Psi_L^{1/q}(Q,\bar{Q})=f_{Q,\bar{Q}}^{1/q}(\{z_i\}) e^{-\sum_i|z_i|^2/2},
\end{align}
where $f_{Q,\bar{Q}}^{1/q}(\{z_i\})$ is a holomorphic function of the dimensionless complex electron coordinates $z_i=(x_i+iy_i)/\sqrt{2}\ell_B$ given by~\cite{qiu2012}
\begin{align}
f_{Q,\bar{Q}}^{1/q}(\{z_i\})=&\prod_i\sqrt{\frac{\lambda}{2\pi}}e^{-\gamma z_i^2/2}\nonumber\\
&\times\prod_{i<j}\left[z_i-z_j+z_0^2\left(\frac{\partial}{\partial z_i}-\frac{\partial}{\partial z_j}\right)\right]^q 1,
\end{align}
where $\lambda$, $\gamma$, and $z_0^2$ are dimensionless parameters. Interpreting Haldane's intrinsic metric as a nematic order parameter via Eq.~(\ref{gab}), we can relate these dimensionless parameters to our complex nematic variational parameter $|Q|e^{i\phi}$ in the following way,
\begin{align}
\lambda&=\sech (|Q|/2),\\
\gamma&=e^{-i\phi}\tanh (|Q|/2),\\
z_0^2&=\frac{1}{2}e^{i\phi}\sinh |Q|.
\end{align}
When the amplitude of nematic order vanishes $|Q|=0$, we have $\lambda=1$, $\gamma=0$, and $z_0^2=0$, and $\Psi_L^{1/q}(Q,\bar{Q})$ reduces to the usual isotropic Laughlin wave function~\cite{laughlin1983}.

\section{Ising anisotropy from finite-size effects}
\label{app:finitesize}

An important consequence of studying an isotropic-to-nematic transition on a torus of finite dimensions $L_x\times L_y$ is the explicit breaking of the continuous $SO(2)$ rotation symmetry to a discrete $C_4$ rotation symmetry. This has the effect of reducing the continuous degeneracy of broken-symmetry nematic states to a discrete two-fold degeneracy. In this Appendix, we derive an effective Hamiltonian [Eq.~(\ref{Heff})] for this 2D ground state manifold. For $L_x=L_y$ and in the absence of tunneling effects the system is effectively an Ising nematic with two degenerate ground states related by a 90$^\circ$ rotation of the nematic director. As we will see, a change of aspect ratio $L_x\neq L_y$ splits the degenerate ground states by a ``Zeeman'' energy $\propto V^{-1/2}\delta$ for small $\delta$ where $L_y/L_x=1+\delta$ and $V=L_xL_y$ is the system volume, and tunneling effects also induce a splitting $\propto\exp(-\textrm{const.}\times V^{1/4})$.

\subsection{Landau-Ginzburg theory of the isotropic-to-nematic transition}

Assuming we are close enough to the transition, we can ignore the coupling to the topological degrees of freedom of the FQH liquid, i.e., the Chern-Simons gauge fields and Laughlin quasiparticles~\cite{maciejko2013}, and we focus on the nematic sector alone. As mentioned previously, the nematic order parameter can be expressed as a complex scalar $Q=|Q|e^{i\phi}$, where $0\leq\phi<2\pi$ corresponds to physically distinct configurations. We denote the complex conjugate by $\bar{Q}=Q^*$, and use a complex notation for spatial derivatives, $\partial=\partial_x+i\partial_y$ and $\bar{\partial}=\partial_x-i\partial_y$. Under an $SO(2)$ spatial rotation by an angle $\theta$, the fields transform as
\begin{align}\label{rotation}
Q\rightarrow e^{2i\theta}Q,\,
\bar{Q}\rightarrow e^{-2i\theta}\bar{Q},\,
\partial\rightarrow e^{i\theta}\partial,\,
\bar{\partial}\rightarrow e^{-i\theta}\bar{\partial},
\end{align}
so an $SO(2)$ invariant Lagrangian should be invariant under this $U(1)$ transformation. In our earlier work~\cite{maciejko2013}, the nematic sector was described by a Lagrangian of the form
\begin{align}\label{L0}
\mathcal{L}_0=i\lambda \bar{Q}\partial_tQ-\mathcal{H}_0,
\end{align}
with a Hamiltonian
\begin{align}\label{H_0}
\mathcal{H}_0=\kappa\partial\bar{Q}\bar{\partial}Q+r\bar{Q}Q+u\bar{Q}\bar{Q}QQ,
\end{align}
which is manifestly invariant under the transformation (\ref{rotation}). This Landau-Ginzburg theory is meant to be an expansion in powers of $Q,\bar{Q}$ and its derivatives. At this order in the expansion, the Lagrangian (\ref{L0}) has the same form as that for an internal degree of freedom (e.g., an XY model), which is reflected in the fact that the Goldstone mode has an isotropic dispersion $\varepsilon(\b{q})\propto|\b{q}|$. In the nematic language, this means the Frank free energy~\cite{ChaikinLubensky} has equal splay and bend coefficients,
\begin{align}\label{FrankFreeEnergy}
\partial\bar{Q}\bar{\partial}Q\propto\left\{(\nabla\cdot\b{n})^2+[\b{n}\times(\nabla\times\b{n})]^2
\right\},
\end{align}
where $\b{n}=(\cos(\phi/2),\sin(\phi/2))$ is the nematic director, which is related to the nematic order parameter by $Q_{ab}=|Q|(2n_an_b-\delta_{ab})$. Note that there is no twist in 2D, $\b{n}\cdot(\nabla\times\b{n})=0$. In writing Eq.~(\ref{FrankFreeEnergy}) we assumed a uniform magnitude $|Q|$ of the nematic order parameter, which is appropriate for a nonlinear sigma model description of the ordered phase.

To see that we are breaking a spatial symmetry and not just an internal symmetry, we need to go to higher orders in the Landau-Ginzburg expansion. The simplest terms that do this (and give different splay and bend coefficients) are cubic terms of the form $Q\nabla Q\nabla Q$~\cite{lubensky1970}. In our complex notation, four such terms are allowed,
\begin{eqnarray}
\mathcal{L}_\textrm{cubic}&=&
c_1\left[Q(\partial\bar{Q})^2+\mathrm{c.c.}\right]
+ic_2\left[Q(\partial\bar{Q})^2-\mathrm{c.c.}\right]\nonumber\\
&&+c_3(Q\bar{\partial}Q\bar{\partial}\bar{Q}+\mathrm{c.c.})
+ic_4(Q\bar{\partial}Q\bar{\partial}\bar{Q}-\mathrm{c.c.}),\nonumber\\
\end{eqnarray}
where $c_1,c_2,c_3,c_4$ are real coefficients. We now imagine that we are in the nematic phase $\langle Q\rangle=Q_0e^{i\phi_0}\neq 0$ and wish to examine the effects of fluctuations. We write $Q(\b{r},t)=(Q_0+\chi(\b{r},t))e^{i(\phi_0+\phi(\b{r},t))}$ and expand in powers of the fluctuations $\chi(\b{r},t)\ll Q_0$, $\phi(\b{r},t)\ll 2\pi$ and their derivatives. The amplitude fluctuations $\chi$ are massive and can be integrated out. The imaginary-time Lagrangian for the Goldstone mode $\phi$ is of the form
\begin{eqnarray}\label{Lag}
\mathcal{L}(\phi)&=&\frac{1}{2}\bigl\{(\partial_\tau\phi)^2+v^2(\nabla\phi)^2+\lambda(\phi_0)[(\partial_x\phi)^2-(\partial_y\phi)^2]\nonumber\\
&&+2\eta(\phi_0)\partial_x\phi\partial_y\phi\bigr\},
\end{eqnarray}
where $v^2\propto Q_0^2$, and
\begin{align}
\lambda(\phi_0)&=A\cos\phi_0-B\sin\phi_0,\\
\eta(\phi_0)&=B\cos\phi_0+A\sin\phi_0,
\end{align}
with $A,B\propto Q_0^3$, hence close to the transition we have $Q_0\ll 1$ and thus $\lambda,\eta\ll v^2$. The Goldstone mode dispersion is
\begin{align}\label{GoldstoneEq}
\varepsilon(\b{q})=\sqrt{v^2\b{q}^2+\lambda(\phi_0)(q_x^2-q_y^2)+2\eta(\phi_0)q_xq_y},
\end{align}
which is clearly anisotropic.

\begin{widetext}

\subsection{Finite-size corrections to the ground-state energy}

We will now use the approach of Fisher~\cite{fisher1989} to calculate the finite-size corrections to the ground-state energy. We consider putting the system on a torus of dimensions $L_x\times L_y$. We define
\begin{align}\label{DeltaE0}
\Delta E_0(L_x,L_y)=E_0(L_x,L_y)-E_0(L_x=\infty,L_y=\infty),
\end{align}
where $E_0$ is the ground-state energy. From Eq.~(\ref{Lag}), we have
\begin{align}
E_0(L_x,L_y)=\frac{1}{2}\int_{-\infty}^\infty\frac{d\omega}{2\pi}\sum_{q_xq_y}\ln\left(\omega^2+v^2\b{q}^2+\lambda(\phi_0)(q_x^2-q_y^2)+2\eta(\phi_0)q_xq_y\right).
\end{align}
In the limit $L_x,L_y\rightarrow\infty$ we have $\sum_{q_xq_y}\rightarrow L_xL_y\int\frac{d^2q}{(2\pi)^2}$ where the integral ranges over the entire momentum-space plane $\mathbb{R}^2$, hence Eq.~(\ref{DeltaE0}) becomes
\begin{align}
\Delta E_0(L_x,L_y)&=\frac{1}{2}\int_{-\infty}^\infty\frac{d\omega}{2\pi}\left(\sum_{q_xq_y}-L_xL_y\int\frac{d^2q}{(2\pi)^2}\right)\ln\left(\omega^2+v^2\b{q}^2+\lambda(\phi_0)(q_x^2-q_y^2)+2\eta(\phi_0)q_xq_y\right)\\
&=\frac{1}{2}\left(\sum_{q_xq_y}-L_xL_y\int\frac{d^2q}{(2\pi)^2}\right)\varepsilon(\b{q}),
\end{align}
with $\varepsilon(\b{q})$ given in Eq.~(\ref{GoldstoneEq}). Using the Poisson summation formula, we can show that~\cite{fisher1989}
\begin{align}
\sum_{q_xq_y}\varepsilon(\b{q})=\sum_{q_x\in\frac{2\pi\mathbb{Z}}{L_x}}\sum_{q_y\in\frac{2\pi\mathbb{Z}}{L_y}}\varepsilon(q_x,q_y)
=\sum_{\ell_x,\ell_y\in\mathbb{Z}}\int_{-\infty}^\infty d\theta_x\int_{-\infty}^\infty d\theta_y\,e^{2\pi i(\ell_x\theta_x+\ell_y\theta_y)}
\varepsilon\left(\frac{2\pi\theta_x}{L_x},\frac{2\pi\theta_y}{L_y}\right),
\end{align}
whereas
\begin{align}
L_xL_y\int\frac{d^2q}{(2\pi)^2}\varepsilon(\b{q})=\int_{-\infty}^\infty d\theta_x\int_{-\infty}^\infty d\theta_y\,
\varepsilon\left(\frac{2\pi\theta_x}{L_x},\frac{2\pi\theta_y}{L_y}\right),
\end{align}
via a simple rescaling $q_x=\frac{2\pi\theta_x}{L_x}$, $q_y=\frac{2\pi\theta_x}{L_y}$. Therefore
\begin{align}\label{DeltaE0sum}
\Delta E_0(L_x,L_y)&=\frac{1}{2}\sum_{\ell_x\neq 0}\sum_{\ell_y\neq 0}\int_{-\infty}^\infty d\theta_x\int_{-\infty}^\infty d\theta_y\,e^{2\pi i(\ell_x\theta_x+\ell_y\theta_y)}
\varepsilon\left(\frac{2\pi\theta_x}{L_x},\frac{2\pi\theta_y}{L_y}\right)\nonumber\\
&=\frac{1}{2}L_xL_y\sum_{\ell_x\neq 0}\sum_{\ell_y\neq 0}\tilde{\varepsilon}(\ell_xL_x,\ell_yL_y),
\end{align}
where $\tilde{\varepsilon}(x,y)$ is the Fourier transform of $\varepsilon(\b{q})$ defined in Eq.~(\ref{GoldstoneEq}). The Fourier integral is difficult to evaluate exactly, but we can expand in powers of the anisotropy $\frac{\lambda}{v^2},\frac{\eta}{v^2}\ll 1$ if we are close enough to the transition. To first order in these quantities, we have
\begin{align}\label{EpsilonFirstOrder}
\tilde{\varepsilon}(r,\varphi)=v\int_0^\infty\frac{dq\,q^2}{2\pi}\int_0^{2\pi}\frac{d\theta}{2\pi}\,e^{iqr\cos(\theta-\varphi)}\left(1+\frac{\lambda(\phi_0)}{2v^2}\cos 2\theta+\frac{\eta(\phi_0)}{2v^2}\sin 2\theta\right)+\mathcal{O}(\lambda^2,\eta^2,\lambda\eta),
\end{align}
where $\b{q}=q(\cos\theta,\sin\theta)$ and $\b{r}=r(\cos\varphi,\sin\varphi)$. Performing the integration over $\theta$ and $q$, we obtain
\begin{align}\label{epsrphi}
\tilde{\varepsilon}(r,\varphi)=-\frac{v}{2\pi r^3}\left(1+\frac{3\lambda(\phi_0)}{2v^2}\cos 2\varphi+\frac{3\eta(\phi_0)}{2v^2}\sin 2\varphi\right),
\end{align}
hence, from Eq.~(\ref{DeltaE0sum}), we have
\begin{align}\label{DE0IsAnis}
\Delta E_0(L_x,L_y)=\Delta E_0^\textrm{is}+\Delta E_0^\textrm{anis}(\phi_0),
\end{align}
where the isotropic contribution $\Delta E_0^\textrm{is}$ comes from the $\phi_0$-independent term in Eq.~(\ref{epsrphi}), and the anisotropic contribution $\Delta E_0^\textrm{anis}(\phi_0)$ is
\begin{align}
\Delta E_0^\textrm{anis}(\phi_0)=-\frac{3L_xL_y}{8\pi v}\sum_{\ell_x\neq 0}\sum_{\ell_y\neq 0}\frac{\lambda(\phi_0)(\ell_x^2L_x^2-\ell_y^2L_y^2)+2\eta(\phi_0)\ell_x\ell_yL_xL_y}{(\ell_x^2L_x^2+\ell_y^2L_y^2)^{5/2}}.
\end{align}
We change variables from $L_x$ and $L_y$ to the total volume $V=L_xL_y$ and aspect ratio $R=L_y/L_x$. The term proportional to $\eta(\phi_0)$ vanishes because the summand is odd in $\ell_x$ or $\ell_y$. We have
\begin{align}
\Delta E_0^\textrm{anis}(\phi_0,V,R)=-\frac{3\lambda(\phi_0)}{8\pi v\sqrt{V}}f(R),
\end{align}
where
\begin{align}
f(R)=R^{3/2}\sum_{\ell_x\neq 0}\sum_{\ell_y\neq 0}\frac{\ell_x^2-R^2\ell_y^2}{(\ell_x^2+R^2\ell_y^2)^{5/2}}.
\end{align}
This function can be evaluated numerically. Alternatively, for aspect ratios close to one, $R=1+\delta$ with $\delta\ll 1$, we can expand for small $\delta$. We find
\begin{align}
f(R)=\frac{\delta}{2}\sum_{\ell_x\neq 0}\sum_{\ell_y\neq 0}\frac{3\ell_x^4-14\ell_x^2\ell_y^2+3\ell_y^4}{(\ell_x^2+\ell_y^2)^{7/2}}+\mathcal{O}(\delta^2)
\simeq -1.37\delta+\mathcal{O}(\delta^2),
\end{align}
hence we obtain
\begin{align}
\Delta E_0^\textrm{anis}(\phi_0,V,\delta)\simeq\textrm{const.}\times\frac{\lambda(\phi_0)\delta}{\sqrt{V}},
\end{align}
for $\delta\ll 1$. However, there should be a four-fold anisotropy for a finite-size system when $\delta=0$. The expansion to first order in $\lambda,\eta$ in Eq.~(\ref{EpsilonFirstOrder}) was insufficient. We go back and expand to second order. The correction to $\tilde{\varepsilon}(\b{r})$ at this order is
\begin{align}
\delta\tilde{\varepsilon}(r,\varphi)&=-\frac{1}{8v^3}\int_0^\infty\frac{dq\,q^2}{2\pi}\int_0^{2\pi}\frac{d\theta}{2\pi}\,e^{iqr\cos(\theta-\varphi)}\left(
\lambda^2(\phi_0)\cos^22\theta+\eta^2(\phi_0)\sin^22\theta
+2\lambda(\phi_0)\eta(\phi_0)\sin 2\theta\cos 2\theta\right)
\nonumber\\
&=\frac{1}{32\pi v^3r^3}\left\{\eta^2(\phi_0)+\lambda^2(\phi_0)
+15\left[(\eta^2(\phi_0)-\lambda^2(\phi_0)]\cos 4\varphi
-2\eta(\phi_0)\lambda(\phi_0)\sin 4\varphi\right]\right\}.
\end{align}
The term proportional to $\eta(\phi_0)\lambda(\phi_0)$ is odd in $\ell_x$ or $\ell_y$ and vanishes upon summation over $\ell_x,\ell_y$. We find that the dependence on the aspect ratio parameter $\delta$ of the contribution of $\delta\tilde{\varepsilon}(\b{r})$ to $\Delta E_0(L_x,L_y)$ begins at order $\delta^2$ only, so we can set $\delta=0$ in the resulting expressions. To leading (zeroth) order in $\delta$ therefore, we find
\begin{align}
\delta\Delta E_0(\phi_0,V,\delta)&=\frac{1}{64\pi v^3\sqrt{V}}\left\{\left[\eta^2(\phi_0)+\lambda^2(\phi_0)\right]
\sum_{\ell_x\neq 0}\sum_{\ell_y\neq 0}\frac{1}{(\ell_x^2+\ell_y^2)^{3/2}}
+15\left[\eta^2(\phi_0)-\lambda^2(\phi_0)\right]
\sum_{\ell_x\neq 0}\sum_{\ell_y\neq 0}\frac{\ell_x^4-6\ell_x^2\ell_y^2+\ell_y^4}{(\ell_x^2+\ell_y^2)^{7/2}}\right\}\nonumber\\
&\simeq\frac{1}{64\pi v^3\sqrt{V}}\left\{4.22
\left[\eta^2(\phi_0)+\lambda^2(\phi_0)\right]
-15\times 1.94\left[\eta^2(\phi_0)-\lambda^2(\phi_0)\right]\right\}\nonumber\\
&=\frac{Q_0^3}{\sqrt{V}}\left(c+a\cos 2\phi_0+b\sin 2\phi_0\right),
\end{align}
where $a,b,c$ are independent of $\phi_0,V,\delta$. The term proportional to $c$ contributes to the uninteresting isotropic part $\Delta E_0^\textrm{is}$. Ignoring the isotropic terms, the finite-size correction to the ground-state energy in the nematic phase is therefore
\begin{align}\label{DeltaE0final}
\Delta E_0^\textrm{anis}(\phi_0,V,\delta)=\frac{Q_0^2}{\sqrt{V}}\left[Q_0(a\cos 2\phi_0+b\sin 2\phi_0)+\delta(a'\cos\phi_0+b'\sin\phi_0)\right]+\mathcal{O}(\delta^2).
\end{align}
The $a,b$ terms describe the finite-size four-fold anisotropy due to the toroidal geometry, while the $a',b'$ terms (proportional to $\delta$) describe the perturbation due to the aspect ratio.

What about terms of cubic and higher order in $\lambda,\eta$ in Eq.~(\ref{EpsilonFirstOrder})? At order $n$ in the expansion, we will have terms of the form $Q_0^{n+1}\times\{\cos 2n\varphi,\sin 2n\varphi\}\times\{\cos n\phi_0,\sin n\phi_0\}$ summed over the square lattice $(\ell_x,\ell_y)\in\mathbb{Z}^2\backslash\{(0,0)\}$. Due to the symmetries of the lattice certain terms will vanish identically when summed over the entire lattice. The mirror plane at $\ell_y=0$ sends $\varphi\rightarrow-\varphi$, hence $\sin 2n\varphi$ vanishes when summed over for all $n$. The mirror plane at $\ell_x=\ell_y$ sends $\varphi\rightarrow\frac{\pi}{2}-\varphi$, under which $\cos 2n\varphi$ changes sign for odd $n$ and stays unchanged for even $n$. Therefore the terms $\cos 2n\varphi$ with even $n$ survive the sum. In general therefore, we will have
\begin{align}
\Delta E_0^\textrm{anis}(\phi_0,V,\delta)&=\frac{1}{\sqrt{V}}\left[\sum_{n=1}^\infty Q_0^{2n+1}\left(a_n\cos 2n\phi_0+b_n\sin 2n\phi_0\right)+\delta\sum_{n=1}^\infty Q_0^{2n}\left(a'_n\cos(2n-1)\phi_0
+b'_n\sin(2n-1)\phi_0\right)\right]\nonumber\\
&\hspace{5mm}+\mathcal{O}(\delta^2),
\end{align}
where $a_n,b_n,a'_n,b'_n$ are constants. We note that $\Delta E_0^\textrm{anis}(\phi_0,V,0)$ is invariant under $\phi_0\rightarrow\phi_0+\pi$ while $\phi_0$ has periodicity $2\pi$, which implies two degenerate minima for square aspect ratio. Sufficiently close to the critical point where $Q_0\ll 1$, we can neglect terms of higher order in $Q_0$ and work with the simpler potential (\ref{DeltaE0final}).
\end{widetext}

\subsection{Zero-mode dynamics and effective Hamiltonian}

The finite-size correction to the ground-state energy (\ref{DeltaE0final}) corresponds to the zero-point energy of the $\b{q}\neq 0$ Goldstone bosons. However, when discussing symmetry breaking in finite systems it is important to consider the effect of the uniform ($\b{q}=0$) mode~\cite{anderson1952}. The lowest-energy $\b{q}\neq 0$ fluctuation has a finite-size gap $\propto\frac{1}{\sqrt{V}}$. On the other hand, as will be seen shortly, the uniform fluctuation $\b{q}=0$ has a finite-size gap $\propto\frac{1}{V}$, which is much smaller than $\frac{1}{\sqrt{V}}$ in the thermodynamic limit $V\rightarrow\infty$. Therefore the $\b{q}\neq 0$ fluctuations are very fast compared to the uniform fluctuation, and it is appropriate to use an approximation \`{a} la Born-Oppenheimer and think of $\phi_0$ in Eq.~(\ref{DeltaE0final}) as a dynamical field $\phi_0=\phi_0(\tau)$. The kinetic term in the action for $\phi_0$ is obtained from Eq.~(\ref{Lag}) with $\phi(\b{r},\tau)=\phi_0(\tau)$,
\begin{align}
S_0[\phi_0]=\frac{1}{2}\int_0^\beta d\tau\int d^2r\,(\partial_\tau\phi_0)^2=\frac{V}{2}\int_0^\beta d\tau\,(\partial_\tau\phi_0)^2,
\end{align}
which corresponds to the first-quantized Hamiltonian of an $SO(2)$ rotor,
\begin{align}
H_0(\phi_0)=-\frac{1}{2V}\frac{\partial^2}{\partial\phi_0^2},
\end{align}
where the wave functions $\Psi(\phi_0)$ must obey the periodic boundary conditions $\Psi(\phi_0)=\Psi(\phi_0+2\pi)$. Using the ansatz $\Psi(\phi_0)\propto e^{iN\phi_0}$ where $N\in\mathbb{Z}$, we obtain the ``tower-of-states'' spectrum $E(N)=\frac{N^2}{2V}$, hence the uniform fluctuation has a finite-size gap $\propto\frac{1}{V}$. The finite-size energy (\ref{DeltaE0final}) is then essentially a potential for $\phi_0$, and we consider the Hamiltonian
\begin{align}\label{Hphi0}
H(\phi_0)=-\frac{1}{2V}\frac{\partial^2}{\partial\phi_0^2}+U(\phi_0),
\end{align}
where
\begin{align}
U(\phi_0)=\frac{Q_0^2}{\sqrt{V}}\bigl[&Q_0(a\cos 2\phi_0+b\sin 2\phi_0)\nonumber\\
&+\delta(a'\cos\phi_0+b'\sin\phi_0)\bigr].
\end{align}
Since $\frac{1}{\sqrt{V}}\gg\frac{1}{V}$ in the thermodynamic limit, we can minimize the potential energy $U(\phi_0)$ first and consider the kinetic term as a perturbation. Since $\delta\ll 1$, we can in fact minimize the $a,b$ terms first and consider the second term in $U(\phi_0)$ as a perturbation. Because $0\leq\phi_0<2\pi$, the $a,b$ terms have two inequivalent minima. Writing
\begin{align}
a\cos 2\phi_0+b\sin 2\phi_0=A\cos 2(\phi_0-f),
\end{align}
with $A^2=a^2+b^2$ (not the same $A$ as before) and $\tan 2f=b/a$, and assuming $A>0$, there are two degenerate minima at
\begin{align}
\phi_0=\frac{\pi}{2}+f,\hspace{5mm}
\phi_0=\frac{3\pi}{2}+f.
\end{align}
These are the two states of our effective Ising nematic, which we can denote by a pseudospin variable $\uparrow,\downarrow$,
\begin{align}
|\uparrow\rangle=\left|\phi_0=\frac{\pi}{2}+f\right\rangle,\hspace{5mm}
|\downarrow\rangle=\left|\phi_0=\frac{3\pi}{2}+f\right\rangle.
\end{align}
Because these are $\phi_0$-eigenstates, the term proportional to $\delta$ in $U(\phi_0)$ is diagonal in the pseudospin basis. Writing
\begin{align}
a'\cos\phi_0+b'\sin\phi_0=B\cos(\phi_0-f'),
\end{align}
where $B^2=(a')^2+(b')^2$ and $\tan 2f'=b'/a'$, the term proportional to $\delta$ becomes a ``Zeeman'' term for the Ising pseudospin,
\begin{align}
\frac{BQ_0^2\delta}{\sqrt{V}}\cos(\phi_0-f')\rightarrow E_Z\tau_z,
\end{align}
where $\tau_z$ is the third Pauli matrix acting in pseudospin space, and
\begin{align}
E_Z=\frac{BQ_0^2\delta}{\sqrt{V}}\sin(f'-f)=\textrm{const.}\times\frac{Q_0^2\delta}{\sqrt{V}},
\end{align}
is the ``Zeeman'' energy. Finally, besides the $\tau_z$ term, even at $\delta=0$ the kinetic term will lift the degeneracy between the two pseudospin states because of tunneling effects, which corresponds to a $\tau_x$ term. We can study these effects via a simple instanton calculation that is equivalent to the WKB approximation. Ignoring the Zeeman term, the action corresponding to the Hamiltonian (\ref{Hphi0}) is
\begin{align}
S[\phi_0]=\int_0^\infty d\tau\left[\frac{V}{2}\left(\frac{d\phi_0}{d\tau}\right)^2+\frac{AQ_0^3}{\sqrt{V}}\cos 2(\phi_0-f)\right],
\end{align}
at zero temperature. The Euler-Lagrange equation is the sine-Gordon equation,
\begin{align}
V\frac{d^2\phi_0}{d\tau^2}=\frac{dU(\phi_0)}{d\phi_0}.
\end{align}
We search for a kink solution with $\phi_0(\tau=-\infty)=\frac{\pi}{2}+f$ and $\phi_0(\tau=+\infty)=\frac{3\pi}{2}+f$. The solution
\begin{align}
\tilde{\phi}_0(\tau)=\pi+f+2\arctan\left(\tanh\sqrt{\frac{AQ_0^3}{V^{3/2}}}(\tau-\tau_0)\right),
\end{align}
where $\tau_0$ is the (arbitrary) position of the kink, satisfies the equation as well as the boundary conditions at $\tau=\pm\infty$. The action for this instanton is
\begin{align}
S_\textrm{inst}=S[\tilde{\phi}_0]&\simeq\frac{4AQ_0^3}{\sqrt{V}}
\int_{-\infty}^\infty d\tau\,\mathrm{sech}^2\left(2\sqrt{\frac{AQ_0^3}{V^{3/2}}}(\tau-\tau_0)\right)\nonumber\\
&=4\sqrt{A}Q_0^{3/2}V^{1/4},
\end{align}
where we have made the approximation of extending the range of integration from $(0,\infty)$ to $(-\infty,\infty)$. Therefore the final effective Hamiltonian for the 2D ground-state manifold is
\begin{align}\label{Heff}
H_\textrm{eff}=\Delta\tau_x+E_Z\tau_z,
\end{align}
where
\begin{align}
E_Z\propto\frac{Q_0^2\delta}{\sqrt{V}},\hspace{5mm}
\Delta\propto e^{-\textrm{const.}\times Q_0^{3/2}V^{1/4}}.
\end{align}

\section{Ising anisotropy from corrections to effective mass theory}
\label{app:anisotropy}

Even for an infinite system, due to the underlying crystal structure of the material hosting the 2DEG continuous rotation symmetry will be broken down to a discrete point group symmetry, with the consequence that the Goldstone mode in the FQH nematic phase will acquire a gap. In this Appendix we show that for a lattice with $C_4$ symmetry, the gap $E_\text{gap}$ of the quasi-Goldstone mode scales like $E_\text{gap}\sim(a/\ell_B)^4$, where $a$ is the lattice constant and $\ell_B$ is the magnetic length.

\subsection{Corrections to effective mass theory}

As discussed in Appendix~\ref{app:finitesize}, nematic order is described by the Landau-Ginzburg Lagrangian (\ref{L0}). This description assumes the 2DEG has a perfect continuous $SO(2)$ rotation symmetry (\ref{rotation}), which is valid if we consider the microscopic first-quantized Hamiltonian
\begin{align}
H=\sum_{i=1}^N\frac{\boldsymbol{\pi}_i^2}{2m^*}+\sum_{i<j}V(\b{r}_i-\b{r}_j),
\end{align}
for a system of $N$ electrons, with $\boldsymbol{\pi}_i=\b{p}_i-e\b{A}(\b{r}_i)$ and $V(\b{r})$ the Coulomb interaction. In general this symmetry will be broken explicitly by lattice effects. At long wavelengths this is captured by corrections to effective mass theory, which for a lattice with $C_4$ symmetry begins at quartic order in the momentum $\boldsymbol{\pi}_i$. (One could also consider anisotropic corrections to the dielectric tensor implicit in $V(\b{r})$.)

For a lattice with $C_4$ symmetry the leading corrections to effective mass theory near the $\Gamma$ point are
\begin{align}\label{DeltaH}
\Delta H=\sum_{i=1}^N\left(\frac{\alpha}{2}\{\pi_{ix}^2,\pi_{iy}^2\}+\beta\left(\pi_{ix}^4+\pi_{iy}^4\right)\right).
\end{align}
The second and first terms could come from nearest-neighbor $t$ and next-nearest-neighbor $t'$ hopping on the square lattice, respectively. The first term has been symmetrized because $\pi_{ix}$ and $\pi_{iy}$ do not commute. Let us assume that $t'\sim t$ for simplicity, and neglect all factors of order one. Then $ta^2\sim \hbar^2/m^*$ and $\alpha,\beta\sim ta^4/\hbar^4$, where $a$ is the lattice constant. Therefore $\alpha,\beta\sim a^2/m^*\hbar^2$.

\subsection{First-order perturbation theory}

Using the variational wave functions presented in Appendix~\ref{app:trialWF}, we wish to compute perturbatively the form of the symmetry-breaking corrections to the Hamiltonian (\ref{H_0}) that originate from (\ref{DeltaH}):
\begin{align}\label{AnisCorr}
\Delta\mathcal{H}(Q,\bar{Q})=\langle\Psi_L^{1/q}(Q,\bar{Q})|\Delta H|\Psi_L^{1/q}(Q,\bar{Q})\rangle+\ldots,
\end{align}
where the first term corresponds to first-order perturbation theory in $\Delta H$ and $\ldots$ to higher orders in perturbation theory. We assume the trial wave functions $|\Psi_L^{1/q}(Q,\bar{Q})\rangle$ are normalized, $\langle\Psi_L^{1/q}(Q,\bar{Q})|\Psi_L^{1/q}(Q,\bar{Q})\rangle=1$. $\Delta\mathcal{H}(Q,\bar{Q})$ should contain terms that are only invariant under rotations by $\theta=n\pi/2$, $n\in\mathbb{Z}$, corresponding to $Q\rightarrow-Q$, $\bar{Q}\rightarrow-\bar{Q}$ for $n$ odd. These terms break the $U(1)$ symmetry of the nematic to $\mathbb{Z}_2$. Listing the most relevant terms first, we expect to have
\begin{align}\label{DH_mass}
\Delta\mathcal{H}(Q,\bar{Q})=c_1(Q^2+\bar{Q}^2)+ic_2(Q^2-\bar{Q}^2)+\ldots,
\end{align}
where $c_1,c_2$ are real constants. In the nematic phase, these terms will correspond to a mass term for the transverse Goldstone mode.

Defining the inter-Landau level (LL) annihilation operator $b_i=\ell_B(\pi_{ix}+i\pi_{iy})/\sqrt{2}\hbar$ and its Hermitian conjugate $b_i^\dag$, where $\ell_B=\sqrt{\hbar/eB}$ is the magnetic length, we obtain
\begin{align}\label{DeltaH2}
\Delta H&=\frac{\hbar^4}{\ell_B^4}\sum_{i=1}^N\biggl[(\alpha+6\beta)b_i^\dag b_i+\left(\frac{\alpha}{2}+3\beta\right)b_i^\dag b_i^\dag b_ib_i\nonumber\\
&\hspace{5mm}+\left(-\frac{\alpha}{4}+\frac{\beta}{2}\right)(b_ib_ib_ib_i+b_i^\dag b_i^\dag b_i^\dag b_i^\dag)\biggr],
\end{align}
plus a constant that we neglect. Note that $b_i,b_i^\dag$ are dimensionless. The expectation value in $\Psi_L^{1/q}(Q,\bar{Q})$ of the first two terms in Eq.~(\ref{DeltaH2}) vanishes, because $b_i$ acts on a LLL wave function. Very explicitly, consider the action of this operator on the trial wave function. Using the fact that $b_j=\frac{1}{2}z_j+\partial/\partial\bar{z}_j$, we have
\begin{align}
b_j\Psi_L^{1/q}(Q,\bar{Q})=&\left(\frac{1}{2}z_j+\frac{\partial}{\partial\bar{z}_j}\right)f_{Q,\bar{Q}}^{1/q}(\{z_i\})e^{-|z_j|^2}\nonumber\\
&\times\prod_{i\neq j}e^{-|z_i|^2/2}\nonumber\\
&=\frac{1}{2}z_j\Psi_L^{1/q}(Q,\bar{Q})+f_{Q,\bar{Q}}^{1/q}(\{z_i\})e^{-|z_j|^2/2}\nonumber\\
&\times\left(-\frac{1}{2}z_j\right)\prod_{i\neq j}e^{-|z_i|^2/2}\nonumber\\
&=0,
\end{align}
using Eq.~(\ref{AnisLaughlin}). For the same reason, we have
\begin{align}
b_ib_ib_ib_i\Psi_L^{1/q}(Q,\bar{Q})=0,
\end{align}
and the term with $b_i^\dag b_i^\dag b_i^\dag b_i^\dag$ will involve terms such as
\begin{align}
&\int\prod_i d^2z_i \Psi_L^{1/q}(Q,\bar{Q})^* b_j^\dag b_j^\dag b_j^\dag b_j^\dag \Psi_L^{1/q}(Q,\bar{Q})\nonumber\\
&=\int\prod_i d^2z_i \left(b_jb_jb_jb_j\Psi_L^{1/q}(Q,\bar{Q})\right)^* \Psi_L^{1/q}(Q,\bar{Q})\nonumber\\
&=0.
\end{align}
Thus $\Delta\mathcal{H}(Q,\bar{Q})$ in Eq.~(\ref{AnisCorr}) vanishes to first order in perturbation theory, and we must go to second order.

\begin{figure}[t]
\includegraphics[width=\columnwidth]{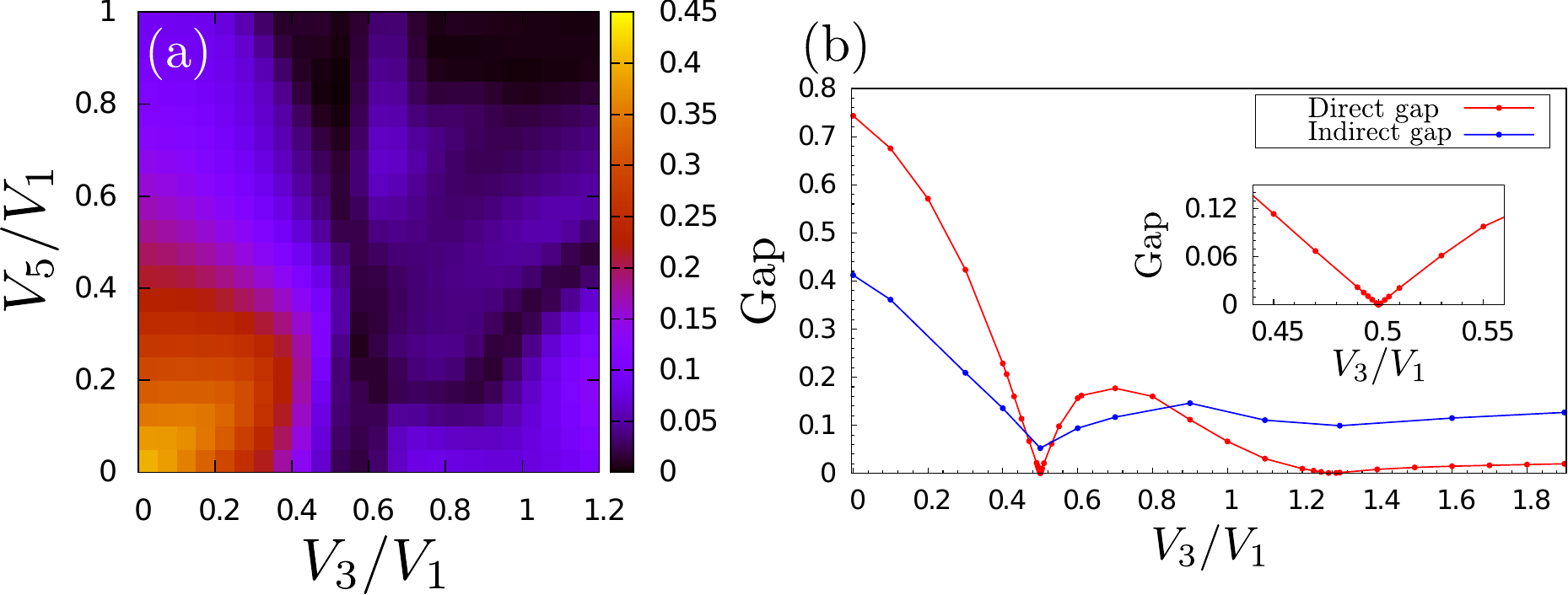}
\caption{(a) Gap between ground and first excited state irrespective of its momentum for the $V_1$-$V_3$-$V_5$ model with $N=11$ electrons. Note that the scale of the gap in Fig.~\ref{fig:fig2}(b) is slightly different but the gross features of the two plots are mostly identical. (b) Gap as a function of the ratio $V_3/V_1$ for $N=13$ electrons and a square aspect ratio. We show both the gap in the zero-momentum sector (red line) and the gap to the first excited with a non zero-momentum (blue line). Note that in order to compare gaps between different values of $V_3/V_1$, we have set the energy scale to be one for the two-particle problem, irrespective of the value of $V_3/V_1$. Inset: Close-up of the gap in the zero-momentum sector around the point $V_3/V_1 \simeq 0.500$ where the gap almost closes ($1.5 \times 10^{-4}$).}\label{fig:IndirectGap}
\end{figure}

\subsection{Second-order perturbation theory}

The second-order correction would look like
\begin{align}
\Delta\mathcal{H}(Q,\bar{Q})=\sum_{n\neq 0}\frac{\langle\Psi_L^{1/q}(Q,\bar{Q})|\Delta H|n\rangle\langle n|\Delta H|\Psi_L^{1/q}(Q,\bar{Q})\rangle}
{E_n-E_0},
\end{align}
where $|n\rangle$, $n\neq 0$ are many-body excited states and $E_0$ is the energy of the trial wave function. We first observe that the terms $b_i^\dag b_i$, $b_i^\dag b_i^\dag b_i b_i$, and $b_ib_ib_ib_i$ will again give a vanishing contribution to $\Delta\mathcal{H}$ since these terms annihilate the trial wave function and thus make the matrix element $\langle n|\Delta H|\Psi_L^{1/q}(Q,\bar{Q})\rangle$ vanish for all $n$. The only nonvanishing contribution must come from the $b_ib_ib_ib_i$ and $b_i^\dag b_i^\dag b_i^\dag b_i^\dag$ terms, which cause LL mixing. Schematically,
\begin{align}
&\Delta\mathcal{H}(Q,\bar{Q})\sim\left(\frac{\hbar^4}{\ell_B^4}\right)^2\left(\frac{a^2}{m^*\hbar^2}\right)^2\nonumber\\
&\times\sum_{n\neq 0}\frac{\langle\Psi_L^{1/q}(Q,\bar{Q})|bbbb|n\rangle\langle n|b^\dag b^\dag b^\dag b^\dag|\Psi_L^{1/q}(Q,\bar{Q})\rangle}{E_n-E_0},
\end{align}
where we used the fact that $\alpha,\beta\sim a^2/m^*\hbar^2$. Now, $|n\rangle$ cannot be in the LLL because this would again lead to the vanishing of the matrix element $\langle\Psi_L^{1/q}(Q,\bar{Q})|bbbb|n\rangle$. The largest contribution to $\Delta\mathcal{H}$ comes from virtual transitions to excited states in the first LL, such that $E_n-E_0\sim\hbar\omega_c$ where $\omega_c=eB/m^*$ is the cyclotron frequency. Since $b,b^\dag$ are dimensionless, the matrix elements are order one. While determining exactly the dependence on $Q,\bar{Q}$ of $\Delta\mathcal{H}(Q,\bar{Q})$ would require a more detailed analysis, we see no reason why the leading terms in Eq.~(\ref{DH_mass}) would not appear. Therefore we expect the magnitude of the coefficients $c_1,c_2$ to be given by
\begin{align}
c_1,c_2\sim\left(\frac{\hbar^4}{\ell_B^4}\right)^2\left(\frac{a^2}{m^*\hbar^2}\right)^2\frac{1}{\hbar\omega_c}\sim\left(\frac{a}{\ell_B}\right)^4\hbar\omega_c,
\end{align}
where we used $\omega_c=eB/m^*$. We thus estimate the gap of the quasi-Goldstone mode in the nematic phase to be
\begin{align}
E_\textrm{gap}\sim Q_0^2\left(\frac{a}{\ell_B}\right)^4\hbar\omega_c,
\end{align}
where $Q_0$ is the (dimensionless) amplitude of the nematic order parameter.

\begin{figure}[t]
\includegraphics[width=0.8\columnwidth]{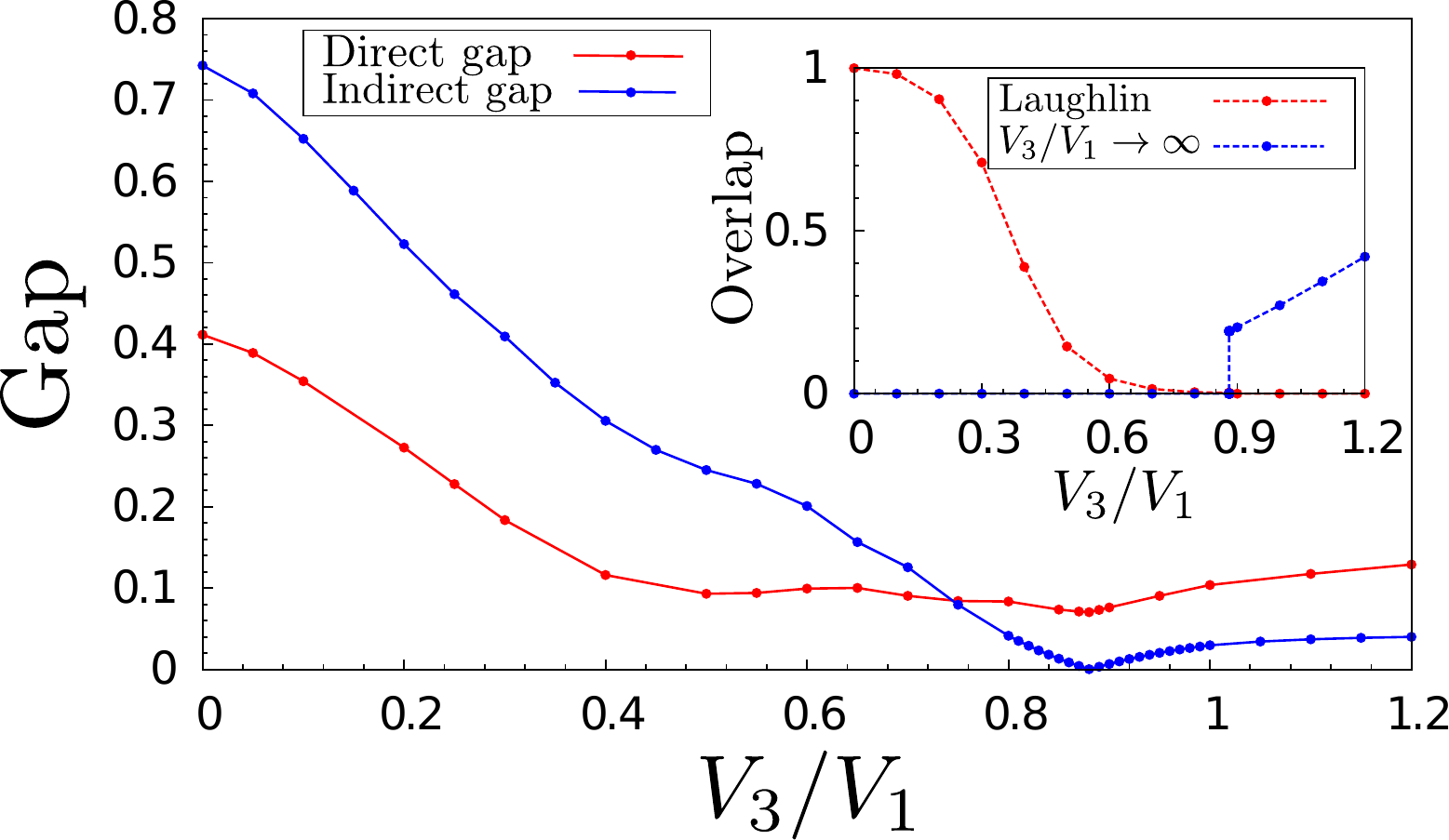}
\caption{Gap as a function of the ratio $V_3/V_1$ for $N=12$ electrons and a square aspect ratio. We show both the gap in the zero-momentum sector (red line) and the gap to the first excited with a non zero-momentum (blue line). In order to compare gaps between different values of $V_3/V_1$, we have set the energy scale to be one for the two-particle problem, irrespective of the value of $V_3/V_1$. Inset: Overlap of the ground state with the $\nu=1/3$ Laughlin state (red dashed line) and the $V_3/V_1\rightarrow\infty$ ground state (blue bashed line) as function of the ratio $V_3/V_1$ for $N=12$ electrons and a square aspect ratio.}\label{fig:V1V3N12}
\end{figure}

In Ref.~\cite{xia2011}, the $\nu=7/3$ plateau is found at $B\approx 2.8$~T, which using the rule of thumb $\ell_B\approx 25\text{ nm}/\sqrt{B[\text{T}]}$ corresponds to a magnetic length $\ell_B\approx 14.9$~nm. The lattice constant of GaAs is $a\approx 5.65$~\AA, which gives
\begin{align}
\left(\frac{a}{\ell_B}\right)^4\approx 2\times 10^{-6},
\end{align}
i.e., a millionth of the cyclotron gap.

\section{Additional numerical results}
\label{app:numerical}

In this Appendix we will provide a more complete numerical survey that might be relevant for the expert readers. In Sec.~\ref{sec:ED}, we have focused on the gap in the zero-momentum sector. We mentioned that looking at the gap to the first excited state (irrespective of its momentum) does not qualitatively change the phase diagram, at least for the system sizes we can reach, i.e., $N=10$ and $N=11$. Indeed, we show the gap to the first excited state for $N=11$ in Fig.~\ref{fig:IndirectGap}(a). It should be compared to Fig.~\ref{fig:fig2}(b) that provides the gap in the zero-momentum sector for this number of fermions. (Performing a similar calculation for $N=13$, including the full $(V_3/V_1, V_5/V_1)$ diagram, would be computationally too demanding.) We have however calculated both the gap in the zero-momentum sector and the indirect gap along the $V_5/V_1=0$ line for this system size [Fig.~\ref{fig:IndirectGap}(b)]. Once again, we see that the gap in the zero-momentum sector is sufficient to characterize the different regions.

\begin{figure}[t]
\includegraphics[width=0.92\columnwidth]{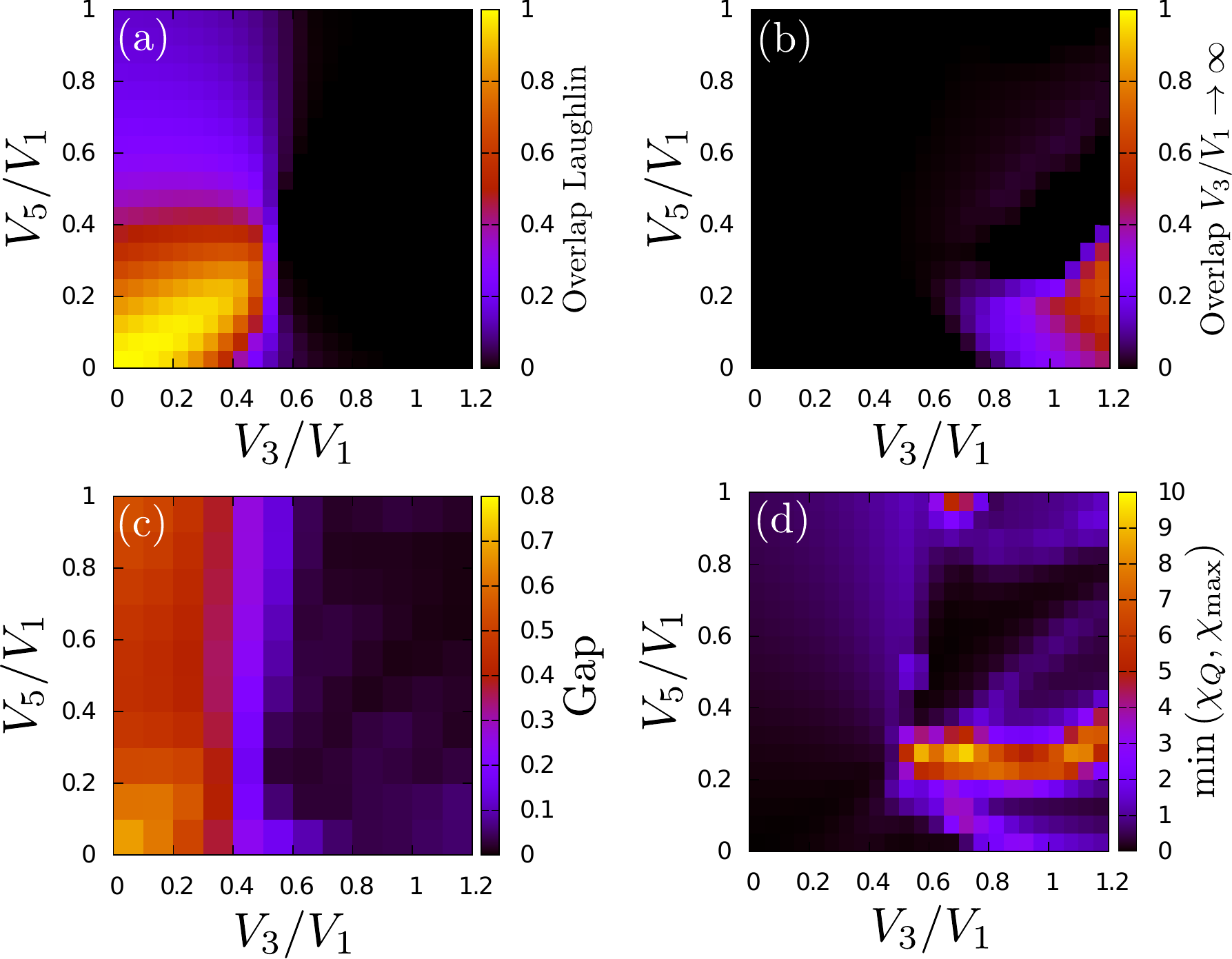}
\caption{Exact diagonalization study of the $V_1$-$V_3$-$V_5$ model with $N=12$ electrons: (a) overlap of the ground state with the $\nu=1/3$ Laughlin state; (b) overlap of the ground state with the $V_3/V_1\rightarrow\infty$ ground state; (c) gap between ground and first excited state in the zero-momentum sector; (d) ground-state nematic susceptibility (\ref{chiQ}), capped for readability at $\chi_\textrm{max}=10$.}\label{fig:DataN12}
\end{figure}

We have mentioned that the system with $N=12$ electrons was not considered in Sec.~\ref{sec:ED} due to an enhanced stability of the Laughlin phase most probably related to commensuration effects. Indeed, on the sphere geometry at the shift of the Laughlin $\nu=1/3$ state, the gap never closes when increasing $V_3/V_1$ for $N=12$ electrons. The torus geometry is not biased by the shift choice. Thus we still observe the gap closing but at a larger value of  $V_3/V_1$ [see Fig.~\ref{fig:V1V3N12}, that should be compared to Fig.~\ref{fig:IndirectGap}(b)]. Despite this greater robustness of the Laughlin phase, the full $(V_3/V_1, V_5/V_1)$ phase diagram for this system size does not exhibit any major qualitative change. Indeed the various quantities (overlaps, gap, and nematic susceptibility) shown in Fig.~\ref{fig:DataN12} for $N=12$ have large similarities with those of Figs.~\ref{fig:fig6} and~\ref{fig:fig2}.

\bibliography{v1v3v5}

\end{document}